%% file: arxiv-tw-Part2-06-23-10.tex
\pdfoutput=1
\documentclass[10pt]{jfm}

\usepackage{psfrag}
\usepackage{graphicx}
\usepackage{amsfonts,amssymb,amsmath}
\usepackage{setspace,subfigure}
\usepackage{latexsym}
\usepackage{pifont}
\usepackage{mathtools}
\usepackage{mathdots}
\usepackage{natbib}

\usepackage{multirow}
\usepackage{multicol}
\usepackage{color}
\usepackage{url}

\input{commands}

\newcommand{\non}{\nonumber}
\newcommand{\ds}{\displaystyle}

\newcommand{\mrd}{\mathrm{d}}

\newcommand{\bu}{{\bf u}}

\newcommand{\bnabla}{\mbox{\boldmath$\nabla$}}

\title{Controlling the onset of turbulence by streamwise traveling waves. \\
Part~2. Direct numerical simulations}

\shorttitle{Controlling the onset of turbulence by traveling waves: DNS study}

\author{Binh K.\ Lieu, Rashad Moarref \and Mihailo R.\ Jovanovi\'c}

\affiliation{Department of Electrical and Computer Engineering,
University of Minnesota, Minneapolis, MN 55455, USA}

\shortauthor{B.\ K.\ Lieu, R.\ Moarref \& M.\ R.\ Jovanovi\'c}

\begin{document}

\maketitle

\begin{abstract}
    This work builds on and confirms the theoretical findings of Part~1 of this paper,~\citet{moajov10}. We use direct numerical simulations of the Navier-Stokes equations to assess the efficacy of blowing and suction in the form of streamwise traveling waves for controlling the onset of turbulence in a channel flow. We highlight the effects of the modified base flow on the dynamics of velocity fluctuations and net power balance. Our simulations verify the theoretical predictions of Part~1 that the upstream traveling waves promote turbulence even when the uncontrolled flow stays laminar. On the other hand, the downstream traveling waves with parameters selected in Part~1 are capable of reducing the fluctuations' kinetic energy, thereby maintaining the laminar flow. In flows driven by a fixed pressure gradient, a positive net efficiency as large as $25 \, \%$ relative to the uncontrolled turbulent flow can be achieved with downstream waves. Furthermore, we show that these waves can also relaminarize fully developed turbulent flows at low Reynolds numbers. We conclude that the theory developed in Part~1 for the linearized flow equations with uncertainty has considerable ability to predict full-scale phenomena.
\end{abstract}

\section{Introduction}
	\label{sec.intro}

The problem of controlling channel flows using strategies that do not require measurement of the flow quantities and disturbances has recently received significant attention. Examples of these sensorless approaches to flow control include wall geometry deformation such as riblets, transverse wall oscillations, and control of conductive fluids using the Lorentz force, to name only a few. \citet{minsunspekim06} used direct numerical simulations (DNS) to
show that surface blowing and suction in the form of an upstream traveling wave (UTW) leads to a sustained sub-laminar drag in a fully developed turbulent channel flow. This motivated~\cite*{marjosmah07} to derive a criterion for achieving sub-laminar drag and to compare laminar and turbulent channel flows with and without control. Furthermore,~\citet{hoefuk09} characterized the mechanism behind UTWs as a {\em pumping\/} rather than as a drag reduction; this is because the UTWs increase flux relative to the uncontrolled flow. Finally,~\citet{bew09} and~\citet*{fuksugkas09} independently established that for any blowing and suction boundary actuation, the power exerted at the walls is always larger than the power saved by reducing drag to sub-laminar levels. This lead the authors of these two papers to conclude that the optimal control solution is to relaminarize the flow.

	Heretofore, sensorless flow control strategies have been designed by combining physical intuition with extensive numerical and experimental studies. For example, a number of simulations on turbulent drag reduction by means of spanwise wall oscillation was conducted by~\citet{quaric04} where $37$ cases of different control parameters were considered. Compared to the turbulent uncontrolled flow, a maximum drag reduction of $44.7 \, \%$ was reported. However, analysis of the power spent by the movement of the walls shows that a maximum net power gain of only $7.3 \, \%$ can be achieved. Even though DNS and experiments offer valuable insight into sensorless strategies, their utility can be significantly enhanced by developing a model-based framework for sensorless flow control design.

	This paper builds directly on the theoretical findings of Part~1,~\citet{moajov10}, where receptivity analysis was used to show that the downstream traveling waves (DTWs) are capable of reducing energy amplification of velocity fluctuations in a transitional channel flow. The effectiveness of DTWs and UTWs in preventing or enhancing transition is examined in this work. In contrast to the current practice, we do not use DNS as a design tool; rather, we utilize them as a means for verification and validation of theoretical predictions offered in Part~1 of this study. Namely, we use DNS to confirm that the DTWs with parameters selected in Part~1 can control the onset of turbulence and achieve positive net efficiency relative to the uncontrolled flow that becomes turbulent. On the contrary, the UTWs enhance transient growth and induce turbulence even when the uncontrolled flow stays laminar. In spite of promoting turbulence, the UTWs with large amplitudes can provide sub-laminar drag coefficient. However, we show that this comes at the expense of poor net power balance in flows driven by a fixed pressure gradient. This is in agreement with~\citet{hoefuk09}, where it was shown that it costs more to achieve the same amount of pumping using wall-transpiration than pressure gradient type of actuation. Our numerical simulations show the predictive power of the theoretical framework developed in Part~1 and suggest that the linearized Navier-Stokes (NS) equations with uncertainty represent an effective control-oriented model for maintaining the laminar flow.

 	Our presentation is organized as follows: in~\S~\ref{sec.numerical}, we present the governing equations, describe the numerical method used in our simulations, and outline the influence of traveling waves on control net efficiency. The evolution of three dimensional (3D) fluctuations around base flows induced by surface blowing and suction is studied in \S~\ref{sec.3Dchannelflow}. We further emphasize how velocity fluctuations affect skin-friction drag coefficient and net power balance. The energy amplification mechanisms are discussed in \S~\ref{sec.mech-energy}, where we show that the DTWs improve transient behavior relative to the uncontrolled flow by reducing the production of kinetic energy. In addition, the effect of traveling waves on coherent flow structures in transitional flows is visualized in~\S~\ref{sec.flow-structures}, where it is shown that the DTWs control the onset of turbulence by weakening the intensity of the streamwise streaks. In~\S~\ref{sec.relaminarization}, we show that the downstream waves designed in Part~1 can also relaminarize fully developed turbulent flows at low Reynolds numbers. We summarize our presentation and give an outlook for future research directions in~\S~\ref{sec.conclusion}.

\section{Problem formulation and numerical method}
	\label{sec.numerical}

   \subsection{Governing equations}
   	\label{sec.goveq}

      \begin{figure}
      	\begin{center}
            \includegraphics[width=0.6\columnwidth]
                {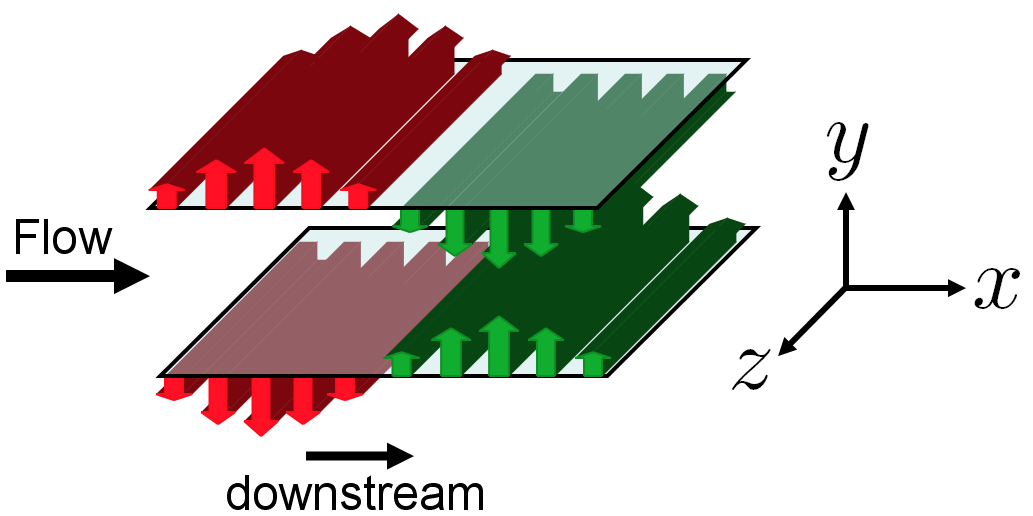}
        	\end{center}
         \caption{A channel flow with blowing and suction along the walls.}
         \label{fig.channel}
      \end{figure}

      We consider a 3D incompressible flow of a viscous Newtonian fluid in a straight channel; see figure~\ref{fig.channel} for geometry. The spatial coordinates $(x, y, z)$ are scaled with the channel half height, $\delta$, and they denote the streamwise, wall-normal, and spanwise directions, respectively; the velocities are scaled with the centerline velocity of the laminar parabolic profile, $U_c$; the pressure is scaled with $\rho U_c^2$, where $\rho$ denotes the fluid density; and the time is scaled with the convective time scale, $\delta/U_c$. The flow is driven by a streamwise pressure gradient and it satisfies the non-dimensional NS and continuity equations
      \beq
         \bu_{t}
         \, = \,
         - \left( \bu \cdot \bnabla \right) \bu
         \, - \,
         \bnabla P
         \, + \,
         ({1}/{R_c}) \Delta \bu,
         ~~
         0
         \, = \,
         \bnabla {\bf \cdot} \bu.
         \label{eq.NScts}
      \eeq
        Here, $R_c$ denotes the Reynolds number, $R_c = U_c \delta / \nu$, $\nu$ is the kinematic viscosity, $\bu$ is the velocity vector, $P$ is the pressure, $\bnabla$ is the gradient, and $\Delta$ is the Laplacian, $\Delta = \bnabla \cdot \bnabla$.

        In addition to the constant pressure gradient, $P_x = -2/R_c$, let the flow be subject to a zero-net-mass-flux surface blowing and suction in the form of a streamwise traveling wave~\citep{minsunspekim06}. The base velocity, $\bu_{b} = (U, V, W = 0)$, represents the steady-state solution to~(\ref{eq.NScts}) in the presence of the following boundary conditions
      \be
         \ba{c}
         V(y \, = \, \pm 1)
         \; = \;
         \mp 2 \, \alpha \,
         \cos \, (\omega_x (x \, - \, c \, t)),
         ~~
         U(\pm 1)
         \; = \;
         V_{y}(\pm 1)
         \; = \;
         W(\pm 1)
         \; = \;
         0,
         \ea
         \label{eq.BCorig}
      \ee
        where $\omega_x$, $c$, and $\alpha$ denote frequency, speed, and amplitude of the streamwise traveling wave. Positive values of $c$ identify a DTW, while negative values of $c$ identify a UTW. In the presence of velocity fluctuations, $\bu$ represents the sum of base velocity, $\bu_{b}$, and velocity fluctuations, $\bv = (u, v, w)$, where $u$, $v$, and $w$ denote the fluctuations in the streamwise, wall-normal, and spanwise directions, respectively.

   \subsection{Numerical method}
		\label{sec.numerical_method}

        The streamwise traveling waves, considered theoretically in Part~1, are tested in DNS of a 3D transitional Poiseuille flow in this work. All DNS calculations are obtained using the code developed by~\cite{gib07}. A multistep semi-implicit Adams-Bashforth/Backward-Differentiation (AB/BDE) scheme described in~\cite{pey02} is used for time discretization. The AB/BDE treats the linear terms implicitly and the nonlinear terms explicitly. A spectral method~\citep{canhusquazan88} is used for the spatial derivatives with Chebyshev polynomial expansion in the wall-normal direction and Fourier series expansion in the streamwise and spanwise directions. Aliasing errors from the evaluation of the nonlinear terms are removed by the 3/2-rule when the horizontal FFTs are computed. We modified the code to account for the streamwise traveling wave boundary conditions~(\ref{eq.BCorig}).

        The NS equations are integrated in time with the objective of computing fluctuations' kinetic energy, skin-friction drag coefficient, and net power balance, \S~\ref{sec.3Dchannelflow}. The velocity field is first initialized with the laminar parabolic profile in the absence of 3D fluctuations, \S~\ref{sec.2Dchannelflow}; this yields the 2D base flow which is induced by the fixed pressure gradient, $P_x = -2/R_c$, and the boundary conditions~(\ref{eq.BCorig}). In simulations of the full 3D flows (cf.\ \S~\ref{sec.3Dchannelflow}), an initial 3D perturbation is superimposed to the base velocity, $\bu_b$. As the initial perturbation, we consider a random velocity field developed by~\cite{gib07} which has the ability to trigger turbulence by exciting all the relevant Fourier and Chebyshev modes. This divergence-free initial condition is composed of random spectral coefficients that decay exponentially and satisfy homogenous Dirichlet boundary conditions at the walls. The flux and energy of the velocity fluctuations are computed at each time step.

        A {\em fixed pressure gradient\/} is enforced in all simulations which are initiated at $R_c = 2000$; this value corresponds to the Reynolds number $R_{\tau} = 63.25$ based on the friction velocity, $u_{\tau}$. Owing to the fixed pressure gradient, the steady-state value of $R_{\tau}$ is the same for all simulations, $R_{\tau} = 63.25$. In addition, we consider a streamwise box length, $L_x = 4 \pi / \omega_x$, for all controlled flow simulations. This box length captures the streamwise modes $k_x = \{ 0, \, \pm \, \omega_x / 2, \, \pm \, \omega_x, \, \pm \, 3 \omega_x / 2, \ldots \}$; relative to Part~1, these modes correspond to the union of the fundamental ($k_x = \{0, \, \pm \, \omega_x, \, \pm 2 \, \omega_x, \, \ldots \}$) and subharmonic ($k_x = \{ \pm \, \omega_x / 2, \, \pm \, 3 \omega_x / 2, \, \ldots \}$) modes. In addition to the uncontrolled flow, we consider three DTWs with $(c = 5,$ $\omega_x = 2,$ $\alpha = \{0.035,0.050,0.125\})$, and three UTWs with $(c = -2,$ $\omega_x = 0.5,$ $\alpha = \{0.015,0.050,0.125\})$. The complete list of the parameters along with the computational domain sizes and the number of spatial grid points is shown in table~\ref{table.dom}. The total integration time is $t_{\mathrm{tot}} = 1000 \, \delta / U_c$. We have verified our simulations by making sure that the changes in results are negligible by increasing the number of wall-normal grid points to $N_y = 97$.

      \begin{table}
         \centering
         \begin{tabular}{ccrrcrrrrr}
            Case\hspace{0.5cm}  & Symbol & \hspace{0.5cm}$c$  & \hspace{0.5cm}$\omega_x$ & \hspace{0.5cm}$\alpha$ & \hspace{0.5cm}$L_x / \delta$ & \hspace{0.5cm}$L_z / \delta$ & \hspace{0.5cm}$N_y$ & \hspace{0.5cm}$N_x$ & \hspace{0.5cm}$N_z$ \\[.1cm] 
            0\hspace{0.1cm} & $\times$ & $-$  & $-$   & \hspace{0.5cm}$-$     & $2 \pi$ & $4 \pi / 3$ & $65$ & $50$  & $50$ \\[.3mm]
            1\hspace{0.1cm} & $\square$ & $5$  & $2$   & \hspace{0.5cm}$0.035$ & $2 \pi$ & $4 \pi / 3$ & $65$ & $50$  & $50$ \\[.3mm]
            2\hspace{0.1cm} & $\circ$ & $5$  & $2$   & \hspace{0.5cm}$0.050$ & $2 \pi$ & $4 \pi / 3$ & $65$ & $50$  & $50$ \\[.3mm]
            3\hspace{0.1cm} & $\lozenge$ & $5$  & $2$   & \hspace{0.5cm}$0.125$ & $2 \pi$ & $4 \pi / 3$ & $65$ & $50$  & $50$ \\[.3mm]
            4\hspace{0.1cm} & $\triangleleft$ & $-2$ & $0.5$ & \hspace{0.5cm}$0.015$ & $8 \pi$ & $4 \pi / 3$ & $65$ & $200$ & $50$ \\[.3mm]
            5\hspace{0.1cm} & $\triangledown$ & $-2$ & $0.5$ & \hspace{0.5cm}$0.050$ & $8 \pi$ & $4 \pi / 3$ & $65$ & $200$ & $50$ \\[.3mm]
            6\hspace{0.1cm} & $\vartriangle$ & $-2$ & $0.5$ & \hspace{0.5cm}$0.125$ & $8 \pi$ & $4 \pi / 3$ & $65$ & $200$ & $50$ \\[.3mm]
         \end{tabular}
         \caption{The computational domain and spatial discretization considered in simulations of the uncontrolled flow, DTWs with $(c = 5,$ $\omega_x = 2,$ $\alpha = \{0.035,0.050,0.125\})$, and UTWs with $(c = -2,$ $\omega_x = 0.5,$ $\alpha = \{0.015,0.050,0.125\})$. Symbols identify the corresponding flow in figures that follow. The box sizes in the streamwise and spanwise directions are denoted by $L_x$ and $L_z$, respectively. The number of grid points in the streamwise, wall-normal, and spanwise directions are represented by $N_i$, $i = \{x, y, z\}$, respectively.}
         \label{table.dom}
      \end{table}

    \subsection{Base flow and nominal net efficiency}
    \label{sec.2Dchannelflow}

\begin{figure}
            \begin{center}
            	\begin{tabular}{cc}
               {\sc downstream:}
                &
                {\sc upstream:}
                \\[-0.15cm]
                \subfigure[]
                {
						\includegraphics[width=0.47\columnwidth]
                {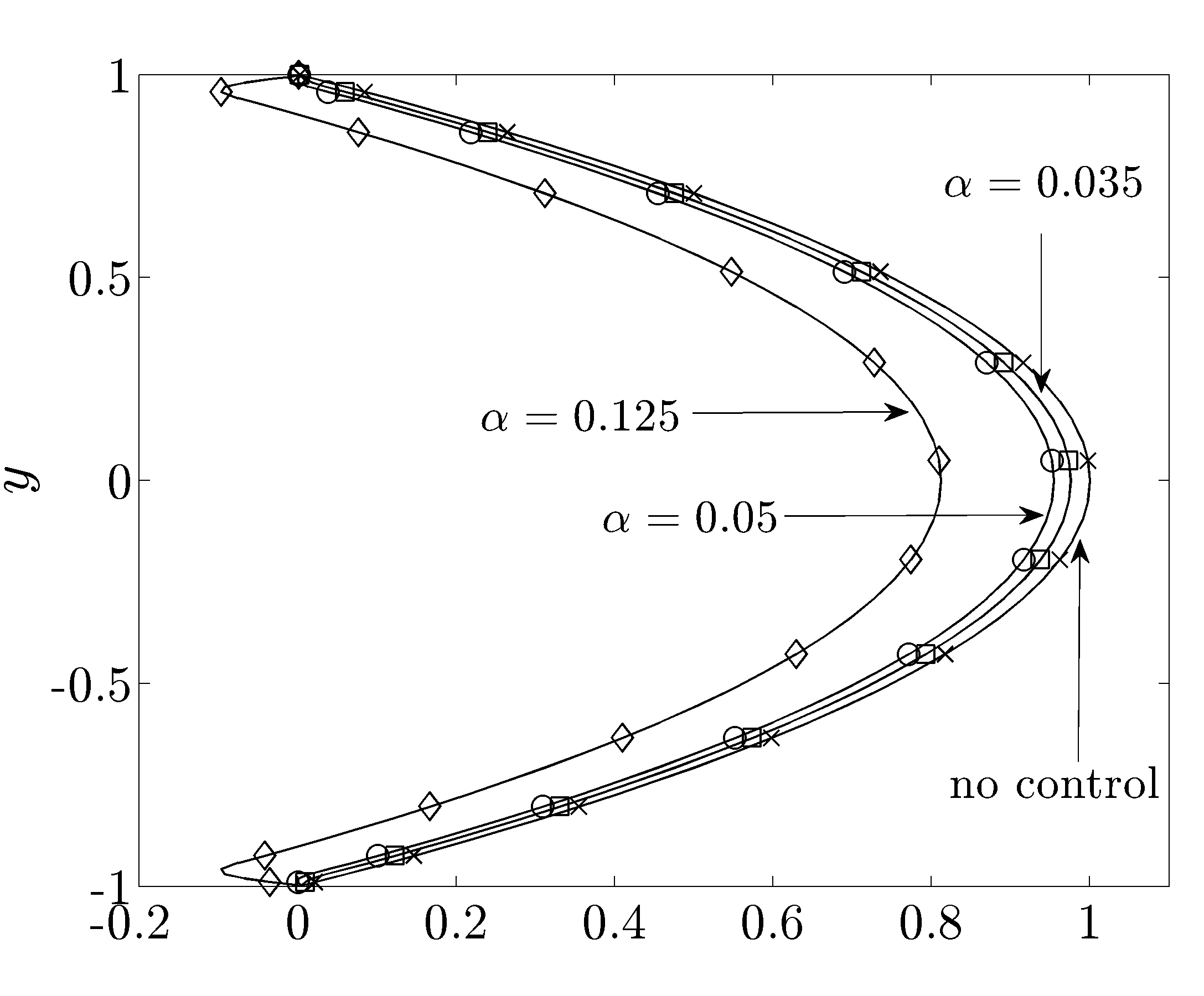}
                \label{fig.umean_dtw}
                }
                &
                \subfigure[]
                {
                \includegraphics[width=0.47\columnwidth]
                {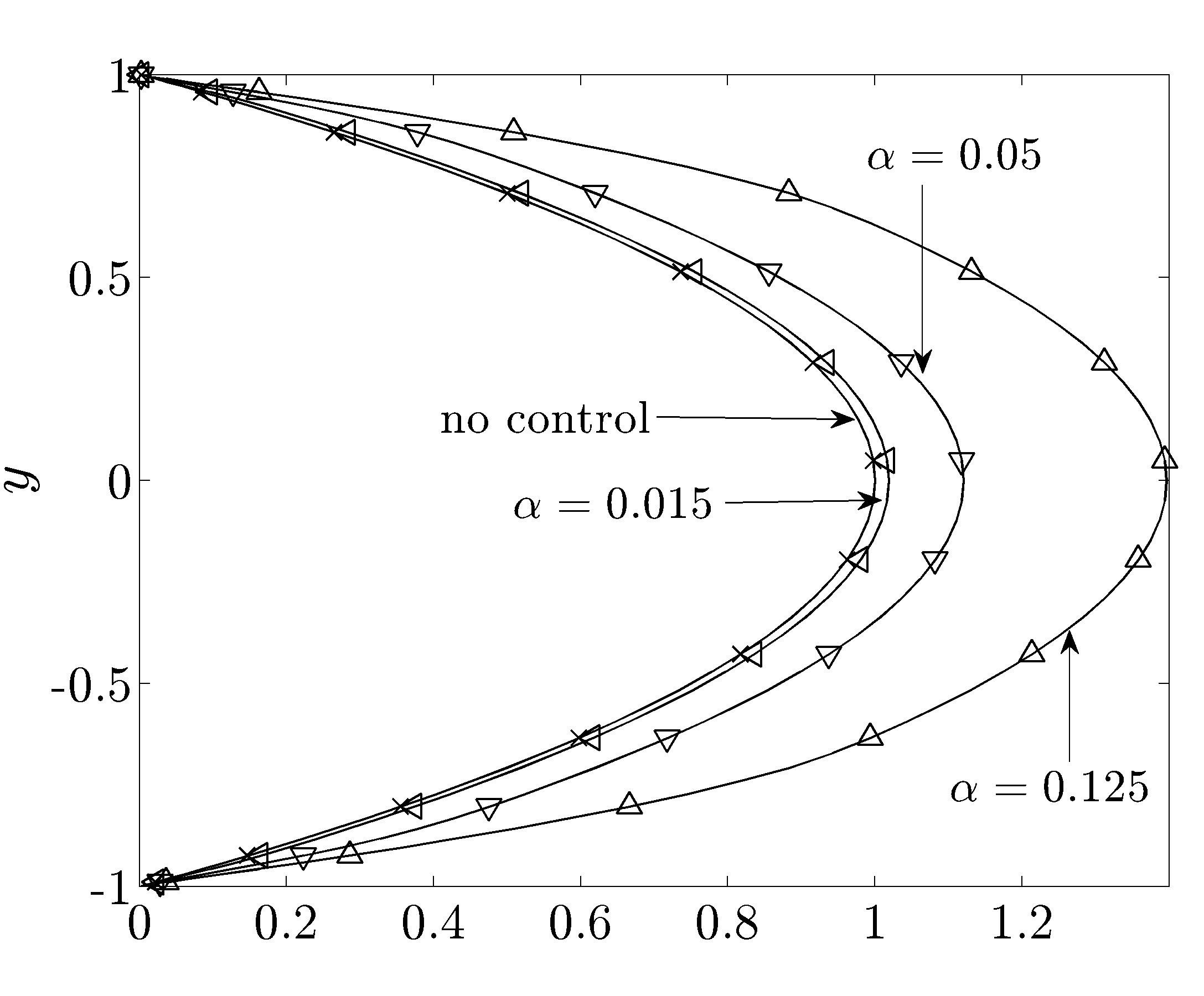}
                \label{fig.umean_utw}
                }
            \end{tabular}
        \end{center}
            \caption{Mean streamwise base velocity, $\overline{U}(y)$, obtained in 2D simulations of the uncontrolled Poiseuille flow with $R_{\tau} = 63.25$, $\times$, and controlled flows subject to: (a) DTWs with $\square$,~$(c = 5$,~$\omega_{x} = 2$,~$\alpha = 0.035)$; $\circ$,~$(c = 5$,~$\omega_{x} = 2$,~$\alpha = 0.05)$; $\lozenge$,~$(c = 5$,~$\omega_{x} = 2$,~$\alpha = 0.125)$; and (b) UTWs with $\triangleleft$,~$(c = -2$,~$\omega_{x} = 0.5$,~$\alpha = 0.015)$; $\triangledown$,~$(c = -2$,~$\omega_{x} = 0.5$,~$\alpha = 0.05)$; $\vartriangle$,~$(c = -2$,~$\omega_{x} = 0.5$,~$\alpha = 0.125)$.}
        \label{fig.umean}
        \end{figure}

        Base velocity, $\bu_{b} = (U(x, y, t), V(x, y, t), 0)$, is computed using DNS of 2D Poiseuille flow with $R_{\tau} = 63.25$ in the presence of streamwise traveling wave boundary control~(\ref{eq.BCorig}). Figure~\ref{fig.umean} shows the mean velocity profiles, $\overline{U}(y)$ (with overline denoting the average over horizontal directions), in uncontrolled flow and in flows subject to selected DTWs and UTWs; these results agree with the results obtained using Newton's method in Part~1. The nominal bulk flux, which quantifies the area under $\overline{U}(y)$,
        \beq
            U_{B,N} \; = \;
            \dfrac{1}{2}
            \ds{\int_{-1}^{1}} \overline{U}(y) \,\mrd y,
            \non
        \eeq
and the nominal skin-friction drag coefficient for three UTWs and three DTWs are reported in table~\ref{table.nominal}. For fixed pressure gradient, $P_x = -2/R_c$, the nominal skin-friction drag coefficient is inversely proportional to square of the nominal flux, i.e.,
        \beq
            C_{f,N}
            \; = \;
            {-2 \, P_x}/{U_{B,N}^2}.
            \label{eq.nom-drag-def}
        \eeq
As shown by~\cite{hoefuk09}, compared to the uncontrolled laminar flow, the nominal flux is reduced (increased) by DTWs (UTWs); according to~(\ref{eq.nom-drag-def}), this results in larger (smaller) nominal drag coefficients, respectively.

        The above results suggest that properly chosen traveling waves can exhibit increased flux compared to the uncontrolled flow. For fixed pressure gradient, this results in production of a driving power
        \be
        \Pi_{prod}
            \; = \;
            -{P_x \, (U_{B,c} \, - \, U_{B,u}) \, (2 L_x L_z)},
            \non
        \ee
where $U_{B,c}$ and $U_{B,u}$ denote the flux of the controlled and uncontrolled flows. The normalized produced power $\% \Pi_{prod}$ is expressed as a percentage of the power spent to drive the uncontrolled flow, $\Pi_u = -P_x \, U_{B,u} \, (2 L_x L_z)$,
        \be
        \% \Pi_{prod}
            \; = \;
            100 \left( U_{B,c} \, - \, U_{B,u} \right)/U_{B,u}.
            \non
        \ee
On the other hand, the input power required for maintaining the traveling waves is obtained from~\citep{cur03}
        \be
            \Pi_{req}
            \; = \;
            \left(
            \left.
            \overline{V P}
            \right|_{y = -1}
            \, - \,
            \left.
            \overline{V P}
            \right|_{y = 1}
            \right)
            \, L_x L_z,
            \non
        \ee
and the normalized required power $\% \Pi_{req}$ is expressed as
        \be
            \% \Pi_{req}
            \; = \;
            100 \, \dfrac{
            \left.
            \overline{V P}
            \right|_{y = -1}
            \, - \,
            \left.
            \overline{V P}
            \right|_{y = 1}}
            {-2 \, P_x \, U_{B,u}}.
            \non
        \ee
In order to assess the efficacy of traveling waves for controlling transitional flows, the control net power is defined as the difference between the produced and required powers~\citep{quaric04}
        \be
            \% \Pi_{net}
            \; = \;
            \% \Pi_{prod}
            \, - \,
            \% \Pi_{req},
            \non
        \ee
where $\% \Pi_{net}$ signifies how much net power is gained (positive $\% \Pi_{net}$) or lost (negative $\% \Pi_{net}$) in the controlled flow as a percentage of the power spent to drive the uncontrolled flow.

        \begin{table}
        \centering
         \begin{tabular}{crrrrrrrr}
             Case\hspace{0.5cm}  & \hspace{0.5cm}$c$  & \hspace{0.5cm}$\omega_x$ & \hspace{1.0cm}$\alpha$ & \hspace{1.0cm}$U_{B,N}$  & \hspace{0.5cm}$10^3 \; C_{f,N}$ & \hspace{0.5cm}$\% \Pi_{prod}$ & \hspace{0.5cm}$\% \Pi_{req}$ & \hspace{0.5cm}$\% \Pi_{net}$\\[.1cm] 
             0\hspace{0.1cm} & $-$  & $-$   & $-$     & $0.6667$ & $4.5000$ & $0$      & $0$      & $0$          \\[.3mm]
             1\hspace{0.1cm} & $5$  & $2$   & $0.035$ & $0.6428$ & $4.8404$ & $-3.58$  & $16.64$  & $-20.22$  \\[.3mm]
             2\hspace{0.1cm} & $5$  & $2$   & $0.050$ & $0.6215$ & $5.1778$ & $-6.77$  & $31.74$  & $-38.51$  \\[.3mm]
             3\hspace{0.1cm} & $5$  & $2$   & $0.125$ & $0.4821$ & $8.6050$ & $-27.69$ & $136.50$ & $-164.19$ \\[.3mm]
             4\hspace{0.1cm} & $-2$ & $0.5$ & $0.015$ & $0.6703$ & $4.4513$ & $2.70$   & $5.46$   & $-2.76$   \\[.3mm]
             5\hspace{0.1cm} & $-2$ & $0.5$ & $0.050$ & $0.7791$ & $3.2949$ & $16.86$  & $37.69$  & $-20.83$  \\[.3mm]

             6\hspace{0.1cm} & $-2$ & $0.5$ & $0.125$ & $1.0133$ & $1.9478$ & $51.99$  & $145.05$ & $-93.06$  \\[.3mm]
        \end{tabular}
        \caption{Nominal results in Poiseuille flow with $R_\tau = 63.25$. The nominal flux, $U_{B,N}$, and skin-friction drag coefficient, $C_{f,N}$, are computed using the base flow described in~\S~\ref{sec.2Dchannelflow}. The produced power, $\% \Pi_{prod}$, required power, $\% \Pi_{req}$, and net power, $\% \Pi_{net}$, are normalized by the power required to drive the uncontrolled flow. The produced and net powers are computed with respect to the laminar uncontrolled flow.}
            \label{table.nominal}
        \end{table}

        The nominal efficiency of the selected streamwise traveling waves in 2D flows, i.e.\ in the absence of velocity fluctuations, is shown in table~\ref{table.nominal}. Note that the nominal net power is negative for all controlled 2D simulations. This is in agreement with a recent study of~\cite{hoefuk09} where it was shown that the net power required to drive a flow with wall transpiration is always larger than in the standard pressure gradient type of actuation.

\section{Avoidance/promotion of turbulence by streamwise traveling waves}
    \label{sec.3Dchannelflow}

    In Part~1 it was shown that a positive net efficiency can be achieved in a situation where the controlled flow stays laminar but the uncontrolled flow becomes turbulent. Whether the controlled flow can remain laminar depends on velocity fluctuations around the modified base flow. In this section, we study the influence of streamwise traveling waves on the dynamics and the control net efficiency. This problem is addressed by simulating a 3D channel flow with initial perturbations which are superimposed on the base velocity induced by the wall actuation. Depending on the kinetic energy of the initial condition, we distinguish three cases: (i) both the uncontrolled and properly designed controlled flows remain laminar ({\em small\/} initial energy); (ii) the uncontrolled flow becomes turbulent, while the controlled flow stays laminar for the appropriate choice of traveling wave parameters ({\em moderate\/} initial energy); and (iii) both the uncontrolled and controlled flows become turbulent for selected traveling wave parameters ({\em large\/} initial energy). Our simulations indicate, however, that {\em poorly\/} designed traveling waves can promote turbulence even for initial conditions for which the uncontrolled flow stays laminar. It was demonstrated in Part~1 that properly designed DTWs are capable of significantly reducing receptivity of velocity fluctuations which makes them well-suited for preventing transition; on the other hand, compared to the uncontrolled flow, the velocity fluctuations around the UTWs at best exhibit similar receptivity to background disturbances. Following Part~1, we present our main results for DTWs with ($c = 5$, $\omega_x = 2$); these results are compared to UTWs with ($c = -2$, $\omega_x = 0.5$) (as selected in~\cite{minsunspekim06}). In both cases, three wave amplitudes are selected (cf.\ table~\ref{table.dom}).

    The 3D simulations, which are summarized in table~\ref{table.3D}, confirm and complement the theoretical predictions of Part~1 at two levels. At the level of controlling the onset of turbulence, we illustrate in~\S~\ref{sec.small} that the UTWs increase receptivity of velocity fluctuations and promote turbulence even for initial perturbations for which the uncontrolled flow stays laminar. In contrast, the DTWs can prevent transition even in the presence of initial conditions with moderate and large energy (cf.\ \S~\ref{sec.moderate} and \S~\ref{sec.large}). At the level of net power efficiency, it is first shown in~\S~\ref{sec.small} that the net power is negative when the uncontrolled flow stays laminar. However, for the uncontrolled flow that becomes turbulent, we demonstrate that the DTWs can result in a positive net efficiency. As discussed in~\S~\ref{sec.moderate} and~\S~\ref{sec.large}, the positive net efficiency is achieved if the required power for maintaining the laminar DTW is less than the produced power. In addition, in~\S~\ref{sec.large}, we highlight an important trade-off that limits the advantages of DTWs in controlling the onset of turbulence in flows subject to large initial conditions. Namely, we show that in this case preventing transition by DTWs requires a large input power that results in a negative efficiency. Our simulations in \S~\ref{sec.moderate} reveal that although UTWs become turbulent, a positive net efficiency can be achieved for small enough wave amplitudes. For the initial conditions with moderate energy, we further point out that the achievable positive net efficiency for UTWs is much smaller than for the DTWs that sustain the laminar flow (cf.\ \S~\ref{sec.moderate}).

    \begin{table}
        \centering
      \begin{tabular}{ccrrrrrrr}
        Initial Energy & Case\hspace{0.5cm}  & \hspace{0.5cm}$c$  & \hspace{0.5cm}$\omega_x$ & \hspace{1.0cm}$\alpha$  & \hspace{0.5cm}$10^3 \; C_f$ & \hspace{0.5cm}$\% \Pi_{prod}$ & \hspace{0.5cm}$\% \Pi_{req}$ & \hspace{0.5cm}$\% \Pi_{net}$\\[.1cm] \hline
         Small      & 0\hspace{0.1cm} &  $-$  &  $-$   &  $-$     & $4.5002$  & $0$      & $0$      & $0$       \\[.3mm]
                    & 2\hspace{0.1cm} &  $5$  &  $2$   &  $0.050$ & $5.1778$  & $-6.77$  & $31.77$  & $-38.54$  \\[.3mm]
                    & 4\hspace{0.1cm} &  $-2$ &  $0.5$ &  $0.015$ & $4.3204$  & $-1.54$  & $5.14$   & $-3.60$   \\[.3mm]
                    & 5\hspace{0.1cm} &  $-2$ &  $0.5$ &  $0.050$ & $5.9426$  & $-16.52$ & $23.22$  & $-39.74$  \\[.3mm]
                    & 6\hspace{0.1cm} &  $-2$ &  $0.5$ &  $0.125$ & $3.6853$  & $12.20$  & $108.41$ & $-96.21$  \\[.5mm] \hline
         Moderate & 0\hspace{0.1cm} &  $-$  &  $-$   &  $-$     & $10.3000$ & $0$      & $0$      & $0$       \\[.3mm]
                    & 1\hspace{0.1cm} &  $5$  &  $2$   &  $0.035$ & $4.9244$  & $52.07$  & $26.44$  & $25.63$   \\[.3mm]
                & 2\hspace{0.1cm} &  $5$  &  $2$   &  $0.050$ & $5.2273$  & $47.35$  & $50.40$  & $-3.05$   \\[.3mm]
                    & 4\hspace{0.1cm} &  $-2$ &  $0.5$ &  $0.015$ & $8.7866$  & $11.36$  & $4.53$   & $6.83$    \\[.3mm]
                    & 5\hspace{0.1cm} &  $-2$ &  $0.5$ &  $0.050$ & $6.7406$  & $31.15$  & $41.96$  & $-10.81$  \\[.3mm]
                    & 6\hspace{0.1cm} &  $-2$ &  $0.5$ &  $0.125$ & $3.9264$  & $77.03$  & $155.80$ & $-78.77$  \\[.5mm] \hline
        Large   & 0\hspace{0.1cm} &  $-$  &  $-$   &  $-$     & $11.2000$ & $0$      & $0$      & $0$       \\[.3mm]
                    & 2\hspace{0.1cm} &  $5$  &  $2$   &  $0.050$ & $11.9000$ & $-3.37$  & $47.90$  & $-51.27$  \\[.3mm]
                    & 3\hspace{0.1cm} &  $5$  &  $2$   &  $0.125$ & $12.1000$ & $-11.31$ & $196.89$ & $-208.20$ \\[.3mm]
                    & 5\hspace{0.1cm} &  $-2$ &  $0.5$ &  $0.050$ & $7.4438$  & $13.68$  & $34.19$  & $-20.51$  \\[.3mm]
                    & 6\hspace{0.1cm} &  $-2$ &  $0.5$ &  $0.125$ & $3.9872$  & $57.75$  & $142.92$ & $-85.17$  \\[.3mm]
      \end{tabular}
      \caption{Results of 3D simulations in Poiseuille flow with $R_\tau = 63.25$ for initial conditions of small, moderate, and large energy (respectively, $E (0) = 2.25 \times 10^{-6}$, $E (0) = 5.0625 \times 10^{-4}$, and $E (0) = 2.5 \times 10^{-3}$). The values of $C_f$, $\% \Pi_{prod}$, $\% \Pi_{req}$, and $\% \Pi_{net}$ correspond to $t = 1000$. For small initial energy, the produced and net powers are computed with respect to laminar uncontrolled flow; for moderate and large initial energies, they are computed with respect to turbulent uncontrolled flow.}
      \label{table.3D}
   \end{table}

   \subsection{Small initial energy}
    \label{sec.small}

        We first consider the initial perturbations with small kinetic energy, $E (0) = 2.25 \times 10^{-6}$, which cannot trigger turbulence in flow with no control. Our simulations show that the DTWs selected in Part~1 of this study improve transient response of the velocity fluctuations; on the contrary, the UTWs considered in~\cite{minsunspekim06} lead to deterioration of the transient response and, consequently, promote turbulence. Since the uncontrolled flow stays laminar, both DTWs and UTWs lead to the negative net efficiency.

\begin{figure}
         \begin{center}
            \begin{tabular}{cc}
               {\sc downstream:}
               &
               {\sc upstream:}
               \\[-0.25cm]
               \subfigure[]
               {
                  \includegraphics[width=0.47\columnwidth]
                  {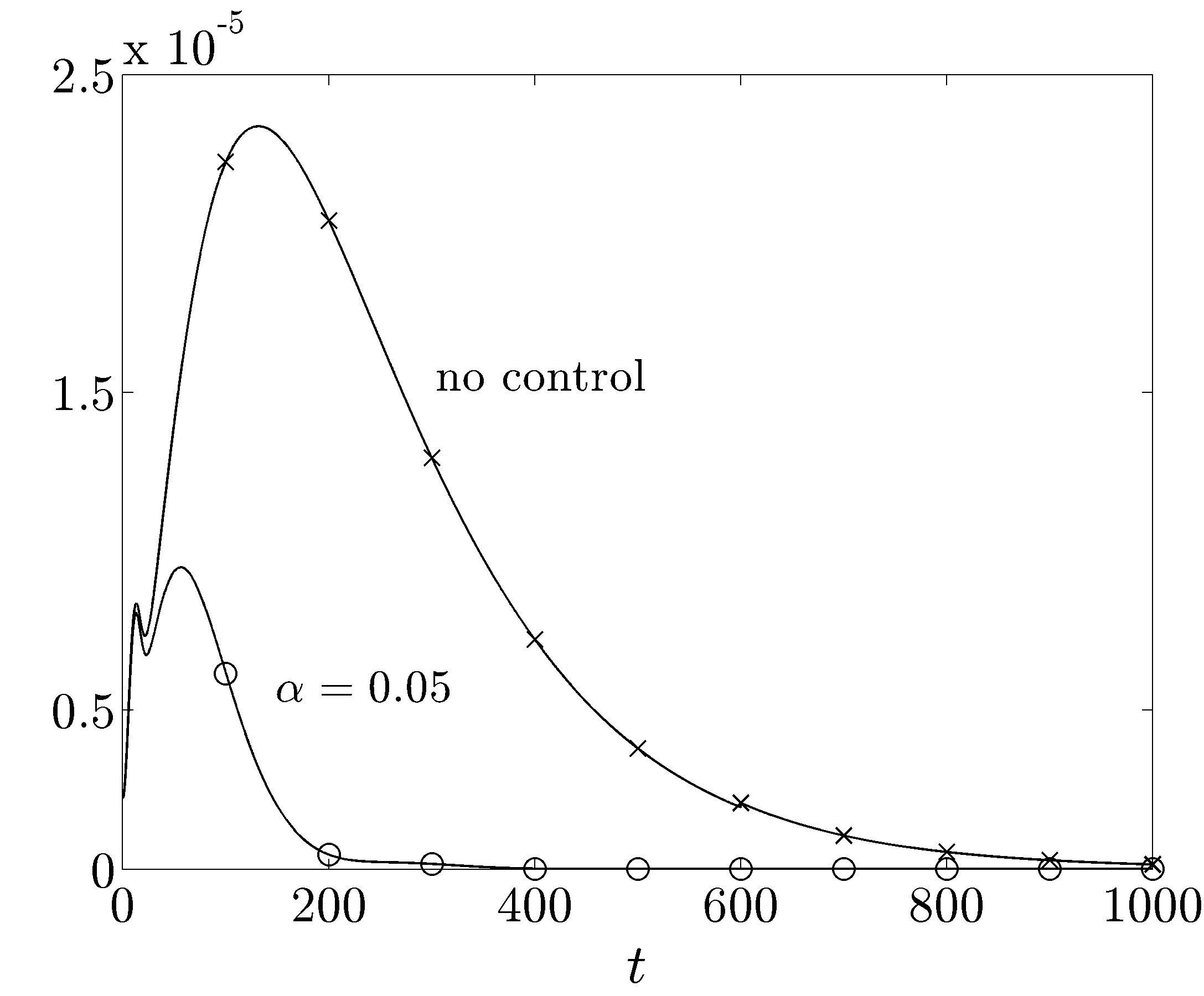}
                  \label{fig.FE-smallIC-dtw}
               }
               &
               \subfigure[]
               {
                  \includegraphics[width=0.47\columnwidth]
                  {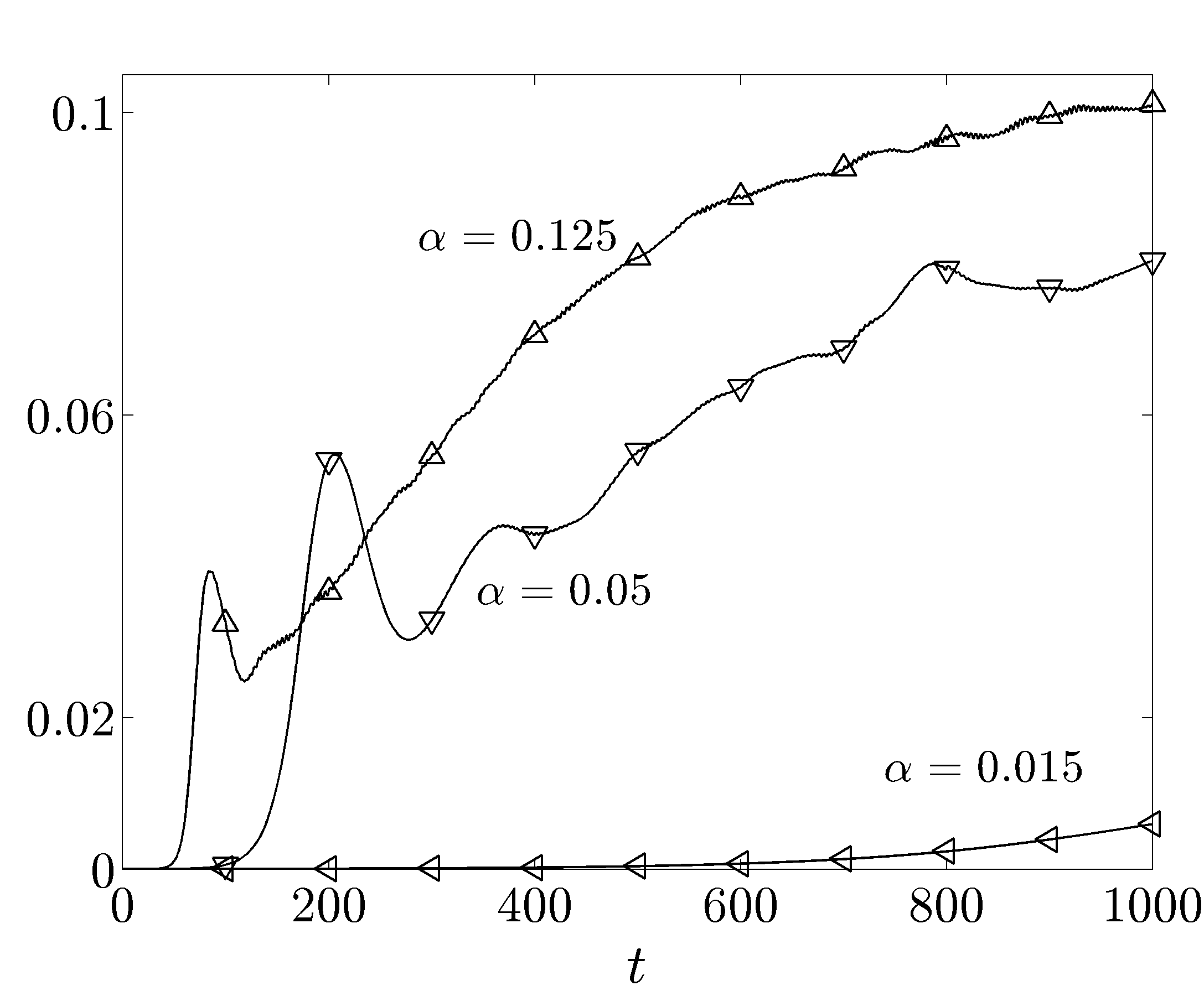}
                  \label{fig.FE-smallIC-utw}
               }
            \end{tabular}
         \end{center}
         \caption{Energy of the velocity fluctuations, $E (t)$, for the initial condition with small energy: (a) $\times$, uncontrolled; $\circ$, a DTW with $(c = 5$, $\omega_{x} = 2$, $\alpha = 0.05)$; and (b) UTWs with $\triangleleft$,~$(c = -2$, $\omega_{x} = 0.5$, $\alpha = 0.015)$; $\triangledown$,~$(c = -2$, $\omega_{x} = 0.5$, $\alpha = 0.05)$; $\vartriangle$,~$(c = -2$, $\omega_{x} = 0.5$, $\alpha = 0.125)$.}
         \label{fig.FE-smallIC}
      \end{figure}

        The energy of velocity fluctuations is given by
      \be
         E(t)
         \; = \;
         \dfrac{1}{\Omega} \,
         \ds{\int_{\Omega}{(u^2 + v^2 + w^2) \, \mrd \Omega}},
         \non
      \ee
where $\Omega = 2 L_x L_z$ is the volume of the computational box. Figure~\ref{fig.FE-smallIC} shows the fluctuations' kinetic energy as a function of time for the uncontrolled flow and controlled flows subject to a DTW with $(c = 5,$ $\omega_x = 2,$ $\alpha = 0.05)$ and three UTWs with $(c = -2,$ $\omega_x = 0.5,$ $\alpha = \{0.015, 0.05, 0.125\})$. As evident from figure~\ref{fig.FE-smallIC-dtw}, the energy of the uncontrolled flow exhibits a transient growth followed by an exponential decay to zero (i.e., to the laminar flow). We see that a DTW moves the transient response peak to a smaller time, which is about half the time at which peak of $E (t)$ in the uncontrolled flow takes place. Furthermore, maximal transient growth of the uncontrolled flow is reduced by approximately $2.5$ times, and a much faster disappearance of the velocity fluctuations is achieved. On the other hand, figure~\ref{fig.FE-smallIC-utw} clearly exhibits the negative influence of the UTWs on a transient response. In particular, the two UTWs with larger amplitudes  significantly increase the energy of velocity fluctuations. We note that the fluctuations' kinetic energy in a flow subject to a UTW with an amplitude as small as $\alpha = 0.015$ at $t = 1000$ is already about two orders of magnitude larger than the maximal transient growth of the flow with no control.

\begin{figure}
         \begin{center}
            \begin{tabular}{cc}
					$C_f$
					&
               $\% \Pi_{req}$
					\\[-0.15cm]
               \subfigure[]
					{
						\includegraphics[width=0.47\columnwidth]
	            	{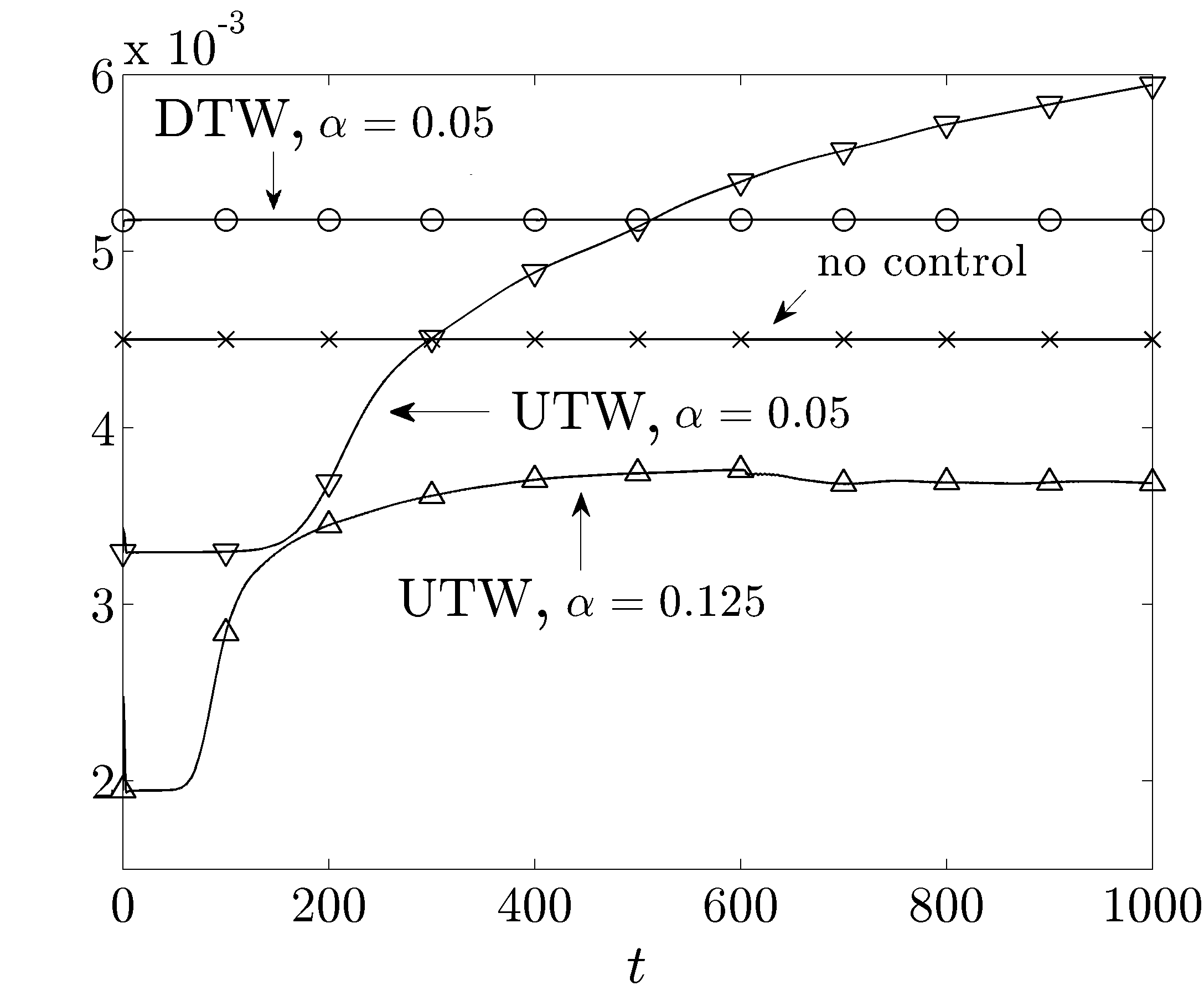}
						\label{fig.Cf-smallIC}
					}
					&
					\subfigure[]
               {
						\includegraphics[width=0.47\columnwidth]
                  {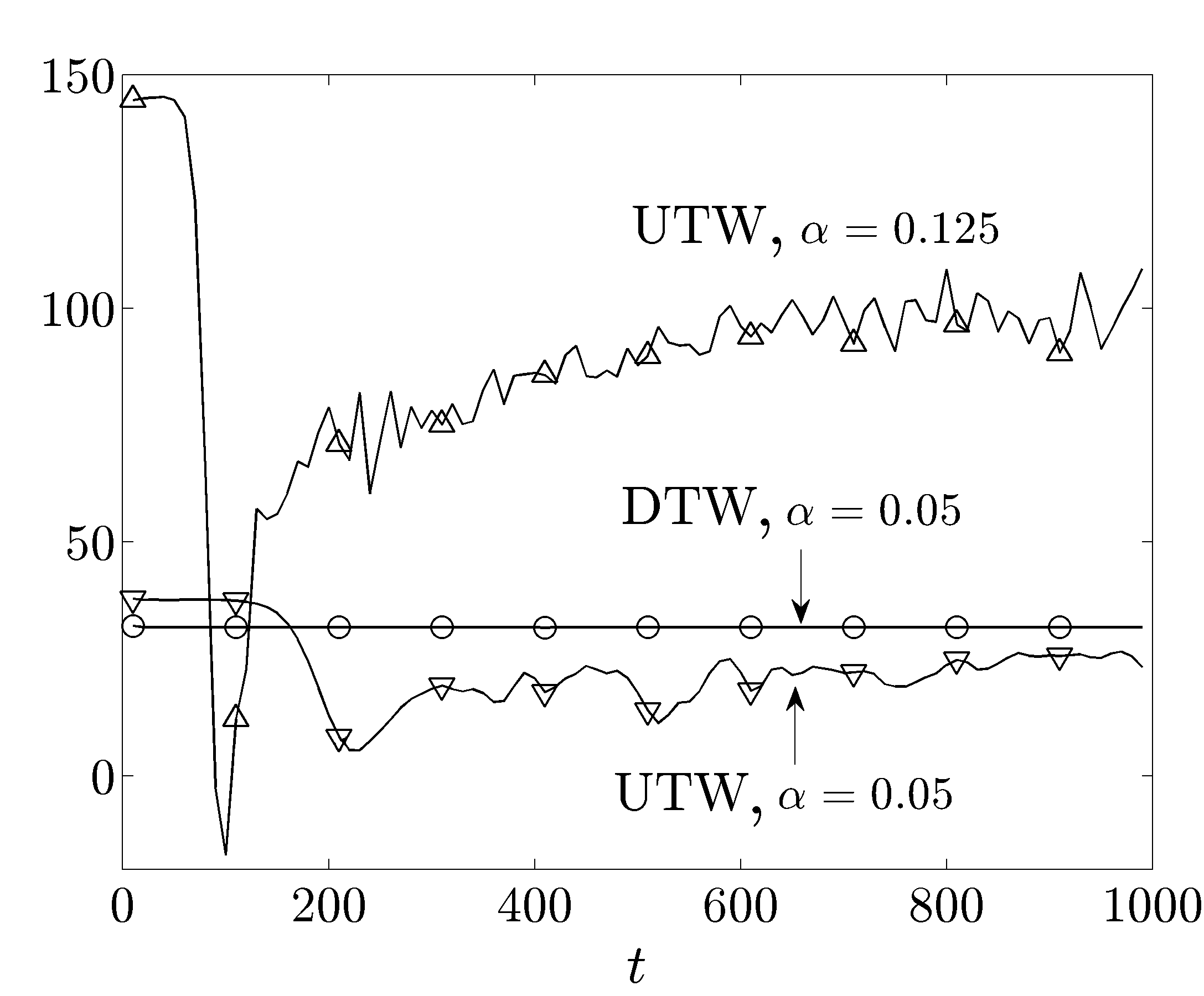}
                  \label{fig.Preq-smallIC}
               }
					\\
               $\% \Pi_{prod}$
					&
					$\% \Pi_{net}$
               \\[-0.15cm]
					\subfigure[]
	            {
	               \includegraphics[width=0.47\columnwidth]
	               {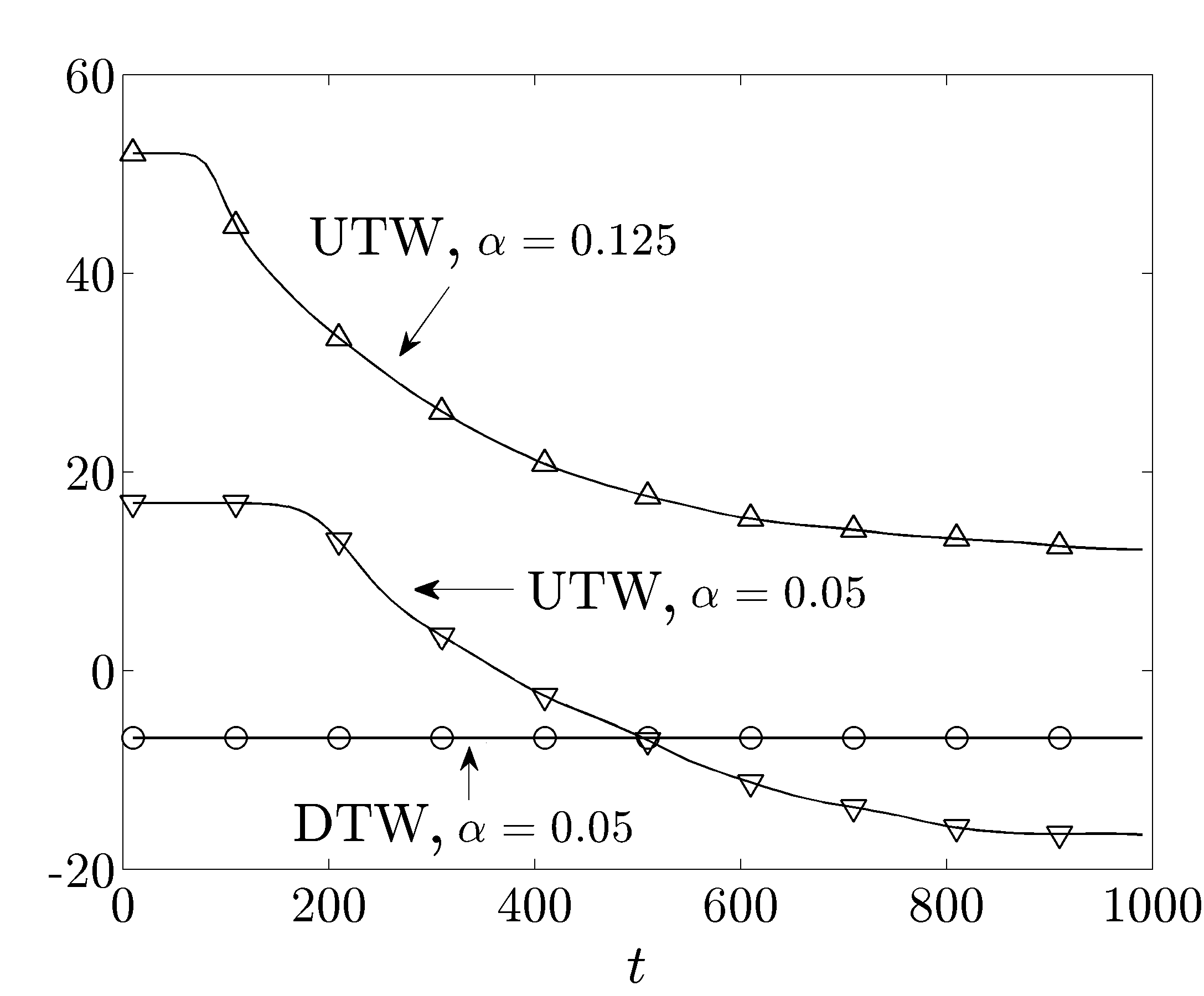}
	               \label{fig.Psav-smallIC}
	            }
					&
					\subfigure[]
               {
						\includegraphics[width=0.47\columnwidth]
            		{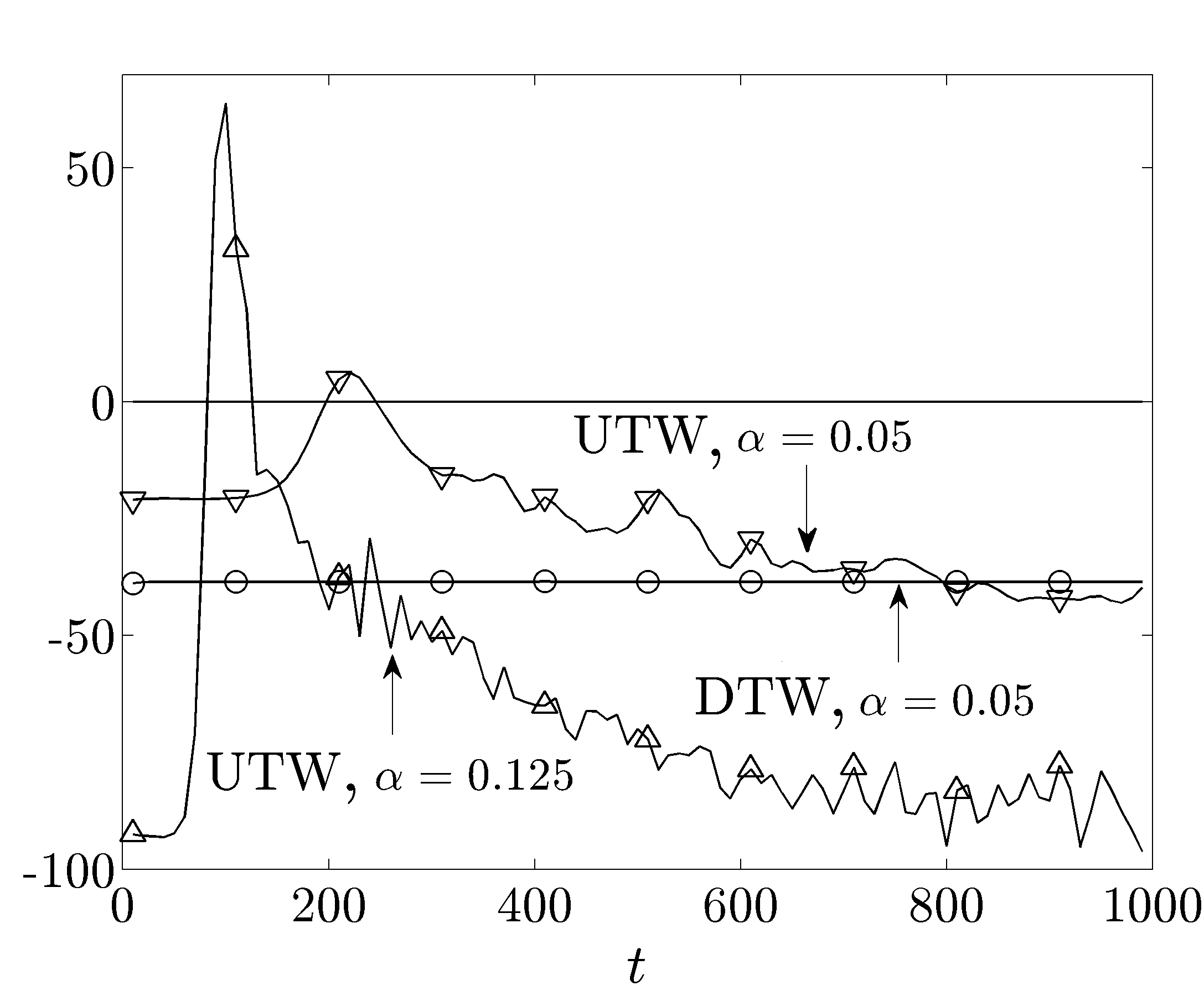}
            		\label{fig.Pnet-smallIC}
               }
				\end{tabular}
         \end{center}
         \caption{(a) Skin-friction drag coefficient, $C_f$; (b) normalized required power, $\% \Pi_{req}$; (c) normalized produced power, $\% \Pi_{prod}$; and (d) normalized net power, $\% \Pi_{net}$, for the initial condition with small energy: $\times$,~uncontrolled flow; $\circ$,~DTW with $(c = 5$, $\omega_{x} = 2$, $\alpha = 0.05)$; and UTWs with $\triangleleft$,~$(c = -2$, $\omega_{x} = 0.5$, $\alpha = 0.015)$; $\triangledown$,~$(c = -2$, $\omega_{x} = 0.5$, $\alpha = 0.05)$; $\vartriangle$,~$(c = -2$, $\omega_{x} = 0.5$, $\alpha = 0.125)$.}
         \label{fig.Cf-Prsn-smallIC}
      \end{figure}

        Figure~\ref{fig.Cf-smallIC} shows the skin-friction drag coefficient,
        \be
        C_f(t)
         \; = \;
         \dfrac{2 \overline{\tau}_w}{U_B^2}
         \; = \;
         \dfrac{1}{R_c\,U_B^2}\,
         \left[
         \left.
         \left(
         \dfrac{\mrd \overline{U}}{\mrd y}
         \, + \,
         \dfrac{\mrd \overline{u}}{\mrd y}
         \right)
         \right|_{y \, = \, -1}
         \, - \,
         \left.
         \left(
         \dfrac{\mrd \overline{U}}{\mrd y}
         \, + \,
         \dfrac{\mrd \overline{u}}{\mrd y}
         \right)
         \right|_{y \, = \, 1}
         \right],
         \non
      \ee
as a function of time for the traveling waves considered in figure~\ref{fig.FE-smallIC}. Here, $\overline{\tau}_w$ denotes the non-dimensional average wall-shear stress and
        \beq
            U_B (t)
            \; = \;
            \dfrac{1}{2}
            \ds{\int_{-1}^{1}}
            \left(\overline{U}(y) \,+\, \overline{u}(y,t) \right) \, \mrd y,
            \non
        \eeq
        is the total bulk flux. Since both the uncontrolled flow and the flow subject to a DTW stay laminar, their steady-state drag coefficients agree with the nominal values computed in the absence of velocity fluctuations (cf.\ tables~\ref{table.nominal} and~\ref{table.3D}). On the other hand, the drag coefficients of the UTWs that become turbulent are about twice the values predicted using the base flow analysis. The large amplification of velocity fluctuations by UTWs is responsible for this increase. The velocity fluctuations in the UTW with $\alpha = 0.015$ are not amplified enough to have a pronounced effect on the drag coefficient. Furthermore, the drag coefficients for the UTWs with $(c = -2$, $\omega_{x} = 0.5$, $\alpha = \{0.05, 0.125\})$ at $t = 1000$ agree with the results of~\cite{minsunspekim06} computed for the fully developed turbulent channel flow. This indicates that the UTWs with larger amplitudes in our simulations have transitioned to turbulence. The above results confirm the theoretical prediction of Part~1 where it is shown that the UTWs are poor candidates for controlling the onset of turbulence for they increase receptivity relative to the uncontrolled flow.

        The normalized required, produced, and net powers for the initial conditions with small kinetic energy are shown in figures~\ref{fig.Preq-smallIC} -~\ref{fig.Pnet-smallIC}. Note that the normalized net power for all traveling waves is negative (cf.\ figure~\ref{fig.Pnet-smallIC}). This confirms the prediction of Part~1 that the net power is negative whenever the uncontrolled flow stays laminar. It is noteworthy that the UTW with $\alpha = 0.125$ has a negative net power despite its significantly smaller drag coefficient compared to the laminar uncontrolled flow. As evident from figures~\ref{fig.Preq-smallIC} and~\ref{fig.Psav-smallIC}, this is because the required power for maintaining this UTW is much larger than the power produced by reducing drag. The above results agree with the studies of~\cite{bew09} and \citet{fuksugkas09} where it was established that the net cost to drive a flow by any transpiration-based strategy is larger than in the uncontrolled laminar flow. Therefore, aiming for sub-laminar drag may not be advantageous from efficiency point of view. Instead, one can design control strategies that yield smaller drag than the uncontrolled turbulent flow and provide positive net power balance (cf.\ \S~\ref{sec.moderate}).

   \subsection{Moderate initial energy}
   	\label{sec.moderate}

      We next consider the velocity fluctuations with moderate initial energy, $E (0) = 5.0625 \times 10^{-4}$. This selection illustrates a situation where the initial conditions are large enough to trigger turbulence in the uncontrolled flow but small enough to allow the properly chosen DTWs to maintain the laminar flow and achieve positive net power balance. As shown in~\S~\ref{sec.small}, the UTWs trigger turbulence even for the initial conditions whose kinetic energy is about $200$ times smaller than the value considered here.
		
		\begin{figure}
        \begin{center}
            \begin{tabular}{cc}
               $E$
               &
               $C_f$
					\\[-0.15cm]
               \subfigure[]
               {
                  \includegraphics[width=0.47\columnwidth]
                  {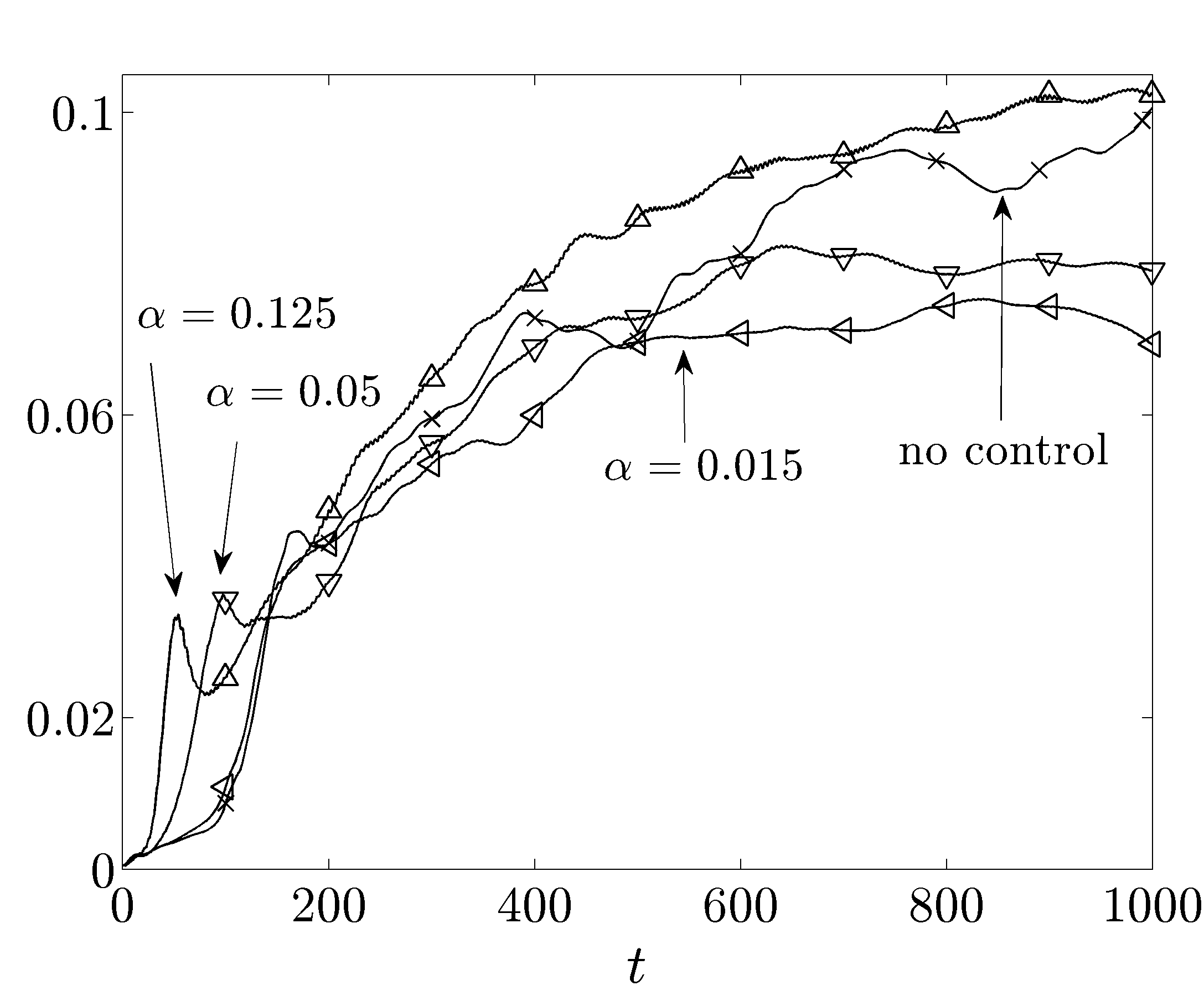}
                  \label{fig.FE-moderateIC-utw}
               }
               &
               \subfigure[]
               {
                  \includegraphics[width=0.47\columnwidth]
                  {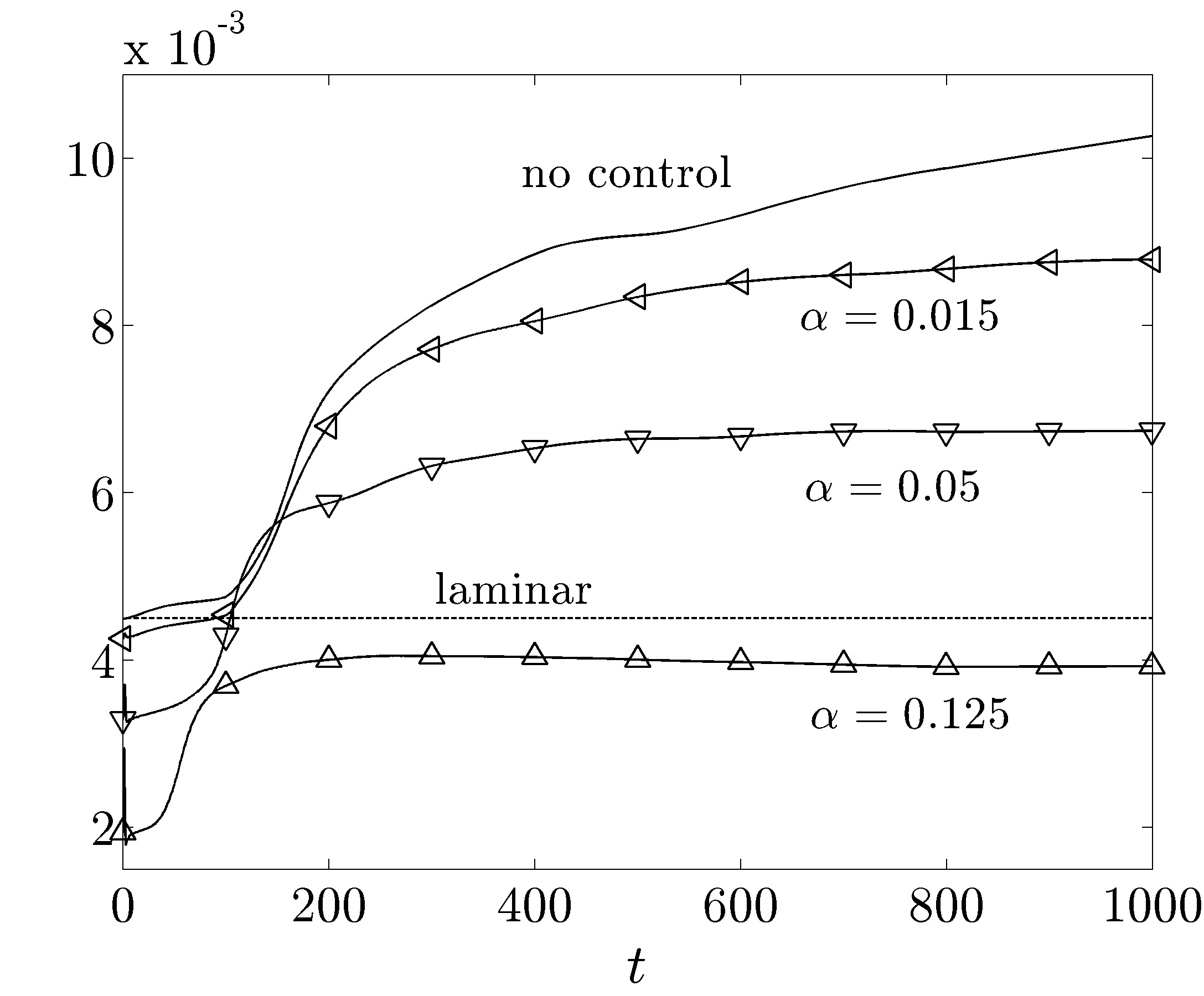}
                  \label{fig.Cf-moderateIC-utw}
               }
					\\
					$\% \Pi_{req}$
					&
					$\% \Pi_{net}$
					\\[-0.15cm]
					\subfigure[]
               {
                  \includegraphics[width=0.47\columnwidth]
                  {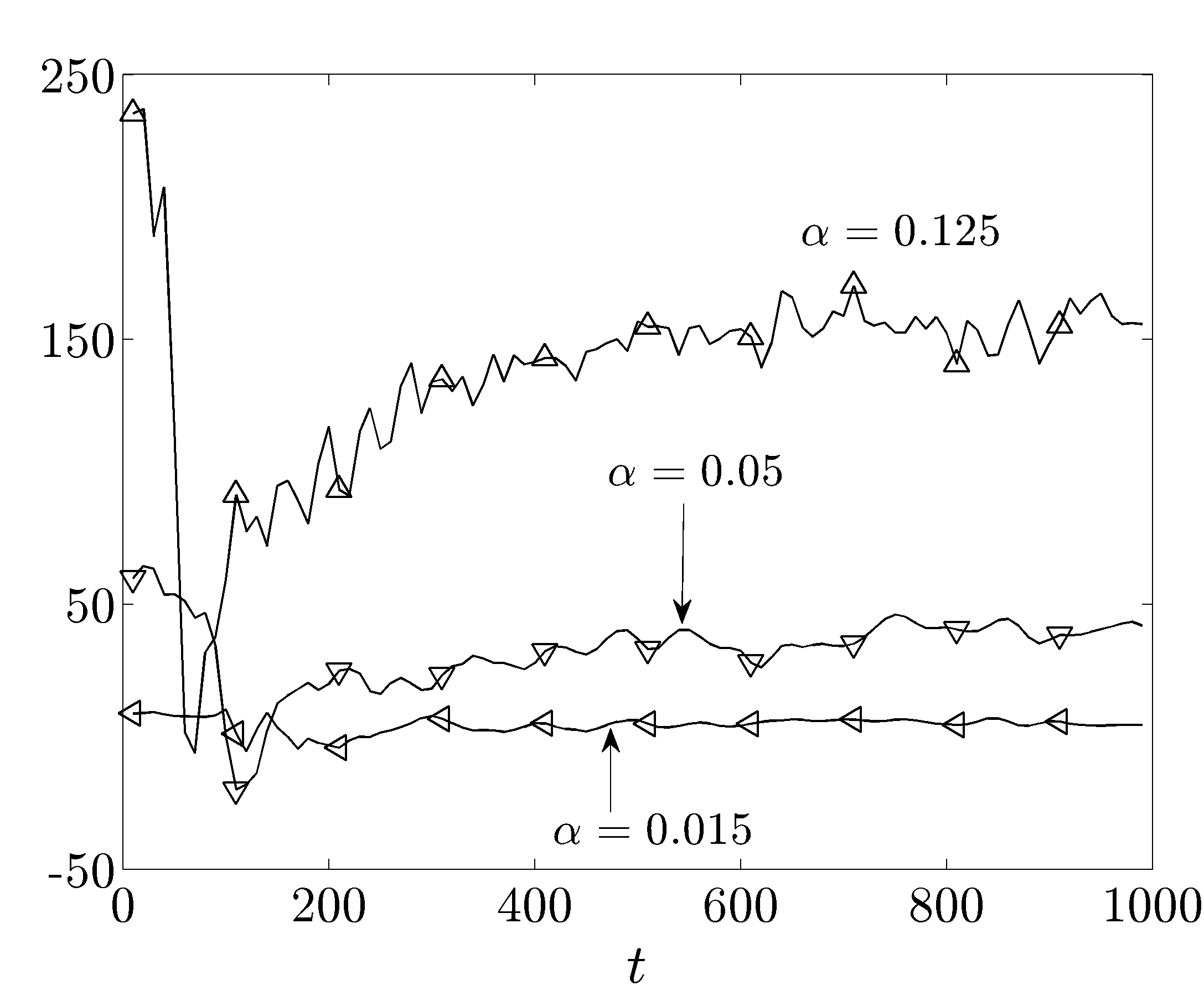}
                  \label{fig.Preq-moderateIC-utw}
               }
					&
					\subfigure[]
               {
                  \includegraphics[width=0.47\columnwidth]
                  {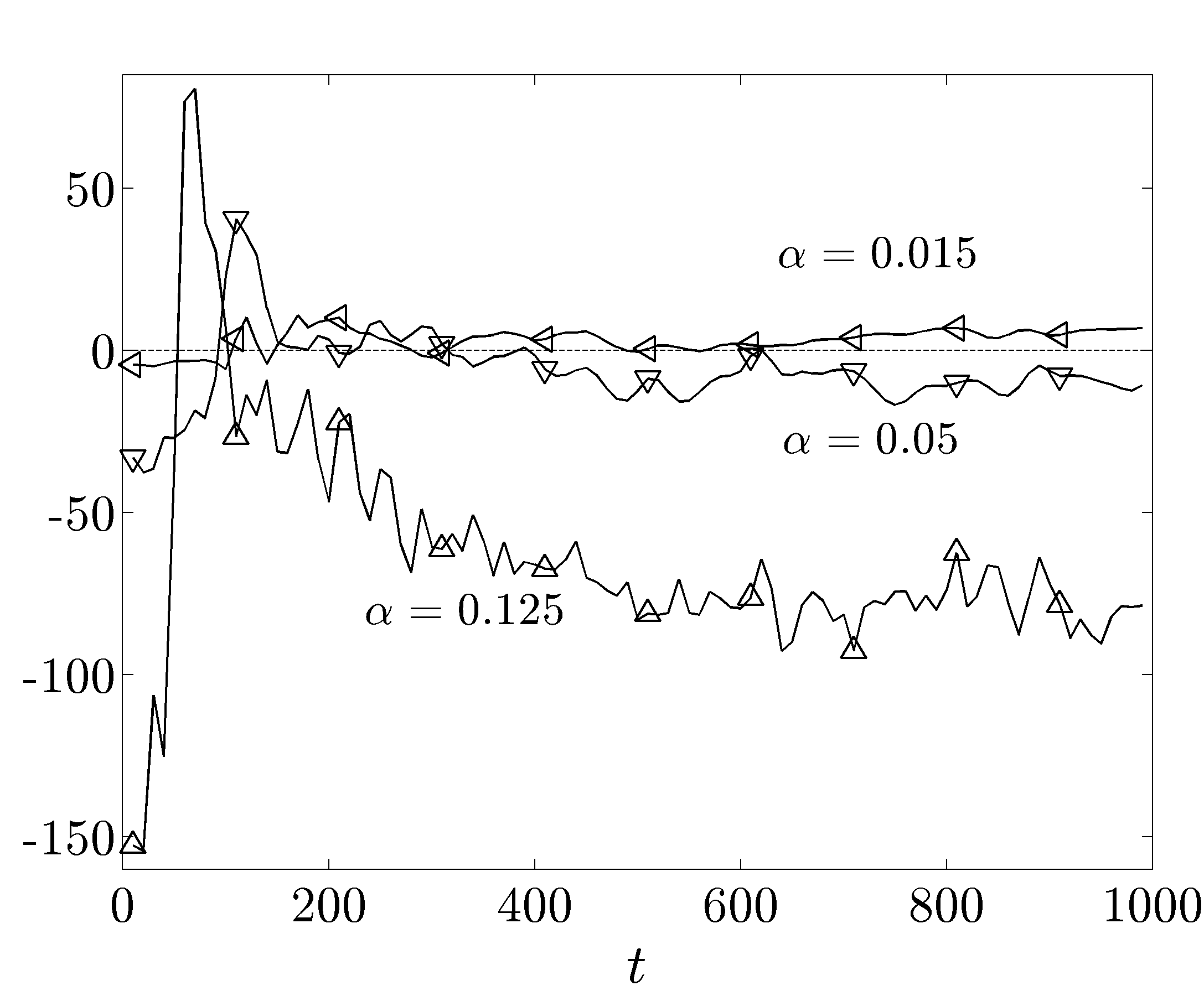}
                  \label{fig.Pnet-moderateIC-utw}
               }
            \end{tabular}
         \end{center}
         \caption{(a) Energy of the velocity fluctuations, $E (t)$; (b) skin-friction drag coefficient, $C_f (t)$; (c) normalized required power, $\% \Pi_{req}$; and (d) normalized net power, $\% \Pi_{net}$, for the initial condition with moderate energy: $\times$, uncontrolled; and UTWs with $\triangleleft$,~$(c = -2, \omega_{x} = 0.5, \alpha = 0.015)$; $\triangledown$,~$(c = -2, \omega_{x} = 0.5, \alpha = 0.05)$; $\vartriangle$,~$(c = -2, \omega_{x} = 0.5, \alpha = 0.125)$.}
         \label{fig.moderateIC-utw}
      \end{figure}

		The energy of the velocity fluctuations and drag coefficients as a function of time for the uncontrolled flow and UTWs are shown in figures~\ref{fig.FE-moderateIC-utw} and \ref{fig.Cf-moderateIC-utw}. Figure~\ref{fig.FE-moderateIC-utw} indicates that the kinetic energy of the uncontrolled flow and the flow subject to UTWs with $(c = -2$,~$\omega_{x} = 0.5$,~$\alpha = \{0.015, 0.05, 0.125\})$ is increased by orders of magnitude which eventually results in transition to turbulence. This large energy amplification of UTWs is captured by the linear analysis around the laminar base flows in Part~1. As evident from figure~\ref{fig.Cf-moderateIC-utw}, the large fluctuations' energy in both the uncontrolled flow and in UTWs yields much larger drag coefficients compared to the nominal values reported in table~\ref{table.nominal}. In addition, figure~\ref{fig.Cf-moderateIC-utw} is in agreement with~\cite{minsunspekim06} where it was shown that the skin-friction drag coefficients of the UTWs are smaller than in the uncontrolled flow that becomes turbulent, and that the UTW with $(c = -2$,~$\omega_{x} = 0.5$,~$\alpha = 0.125)$ achieves a sub-laminar drag. The normalized required and net powers for the UTWs are shown in figures~\ref{fig.Preq-moderateIC-utw} and \ref{fig.Pnet-moderateIC-utw}. Note that the required power for maintaining the UTW with $\alpha = 0.125$ (which yields sub-laminar drag) is so large that it results in a negative net power balance (cf.\ figure~\ref{fig.Pnet-moderateIC-utw}). On the other hand, the UTW with $\alpha = 0.015$ is capable of producing a small positive net power for two main reasons: (i) it has a smaller drag coefficient than the uncontrolled turbulent flow (although it becomes turbulent itself); and (ii) it requires a much smaller power compared to the UTW with $\alpha = 0.125$.
		
		\begin{figure}
        \begin{center}
            \begin{tabular}{cc}
               $E$
               &
               $C_f$
					\\[-0.15cm]
               \subfigure[]
               {
                  \includegraphics[width=0.47\columnwidth]
                  {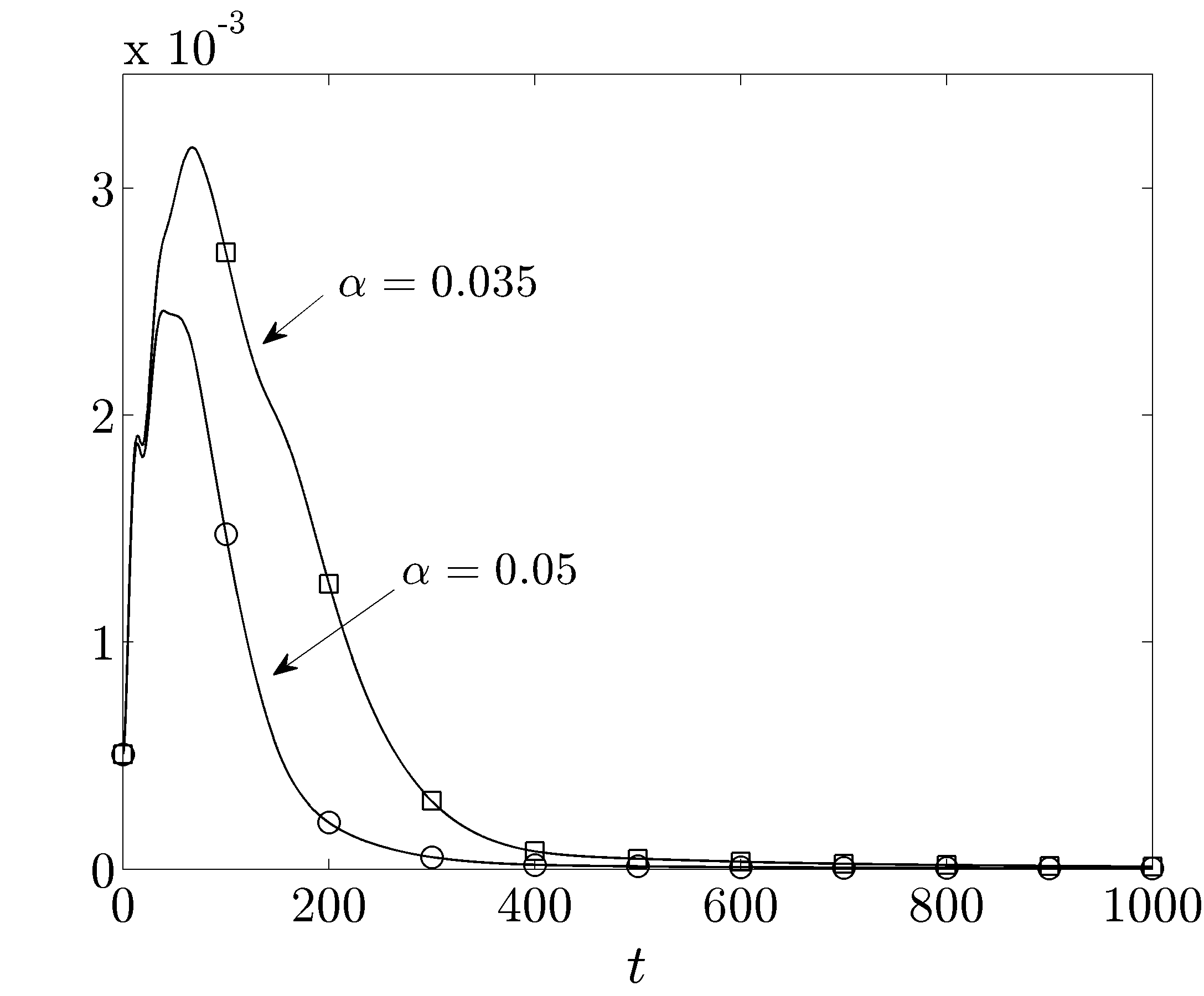}
                  \label{fig.FE-moderateIC-dtw}
               }
               &
               \subfigure[]
               {
                  \includegraphics[width=0.47\columnwidth]
                  {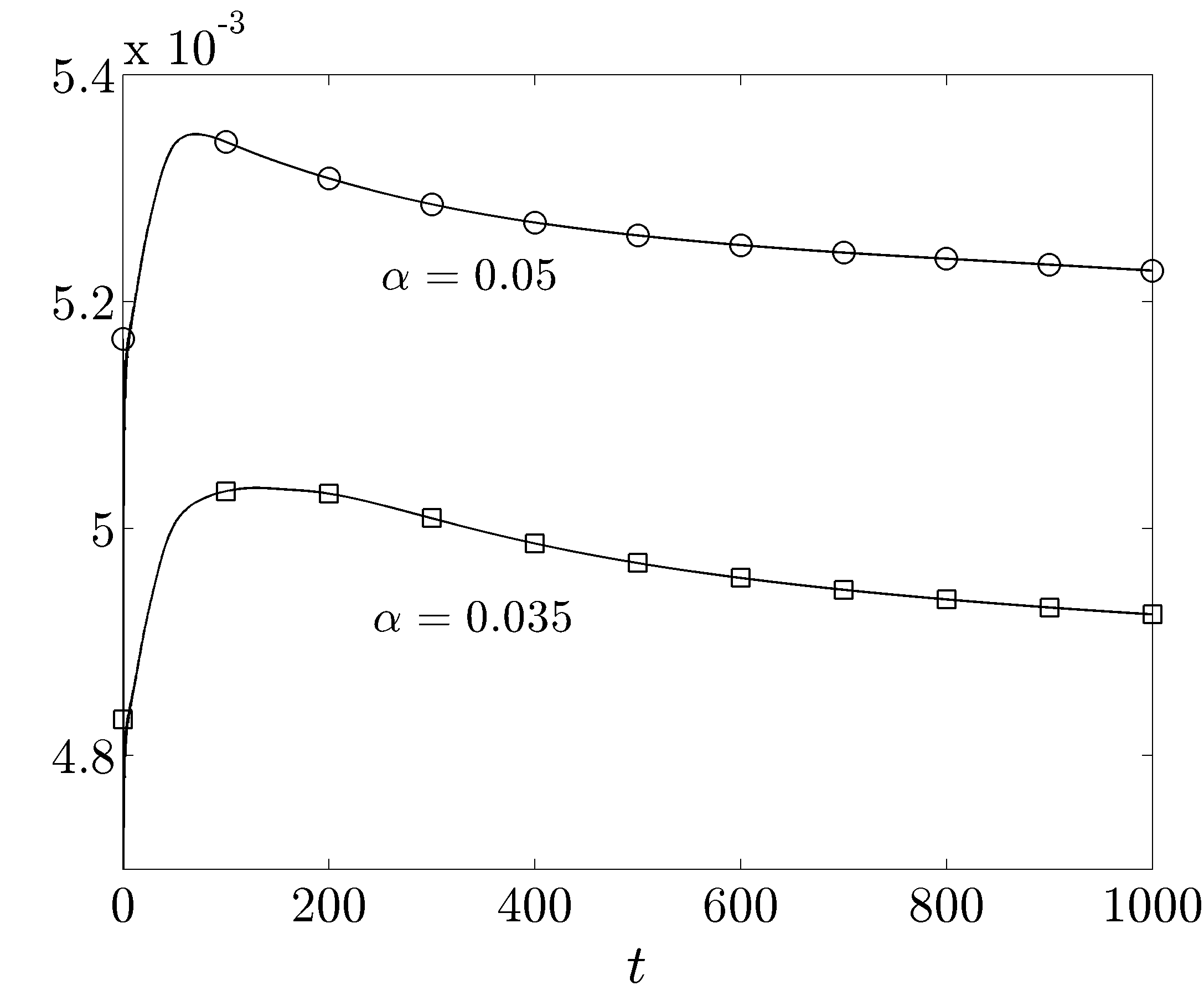}
                  \label{fig.Cf-moderateIC-dtw}
               }
					\\
					$\% \Pi_{req}$, $\% \Pi_{prod}$
					&
					$\% \Pi_{net}$
					\\[-0.15cm]
					\subfigure[]
               {
                  \includegraphics[width=0.47\columnwidth]
                  {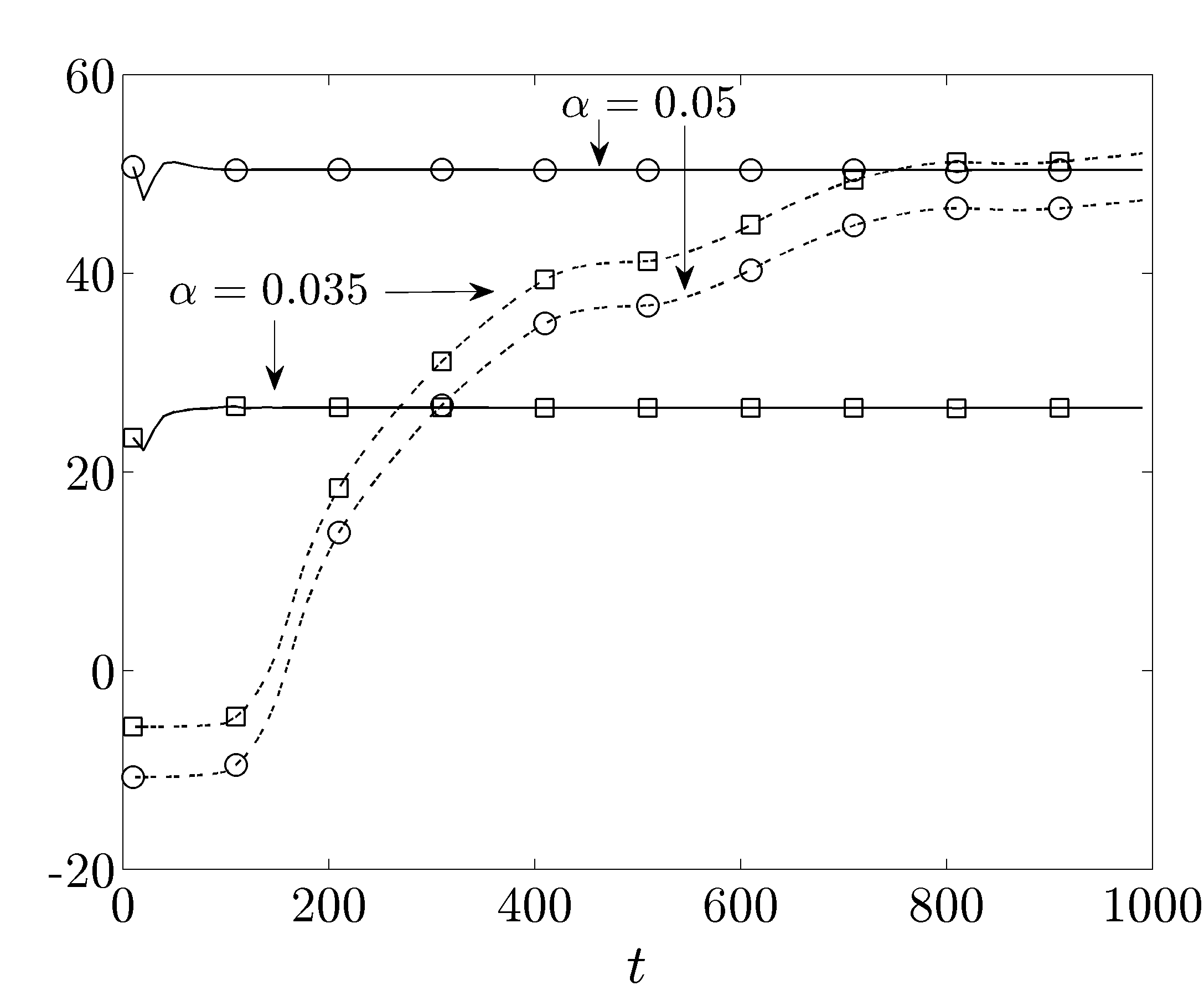}
                  \label{fig.Preq-Psav-moderateIC-dtw}
               }
					&
					\subfigure[]
               {
                  \includegraphics[width=0.47\columnwidth]
                  {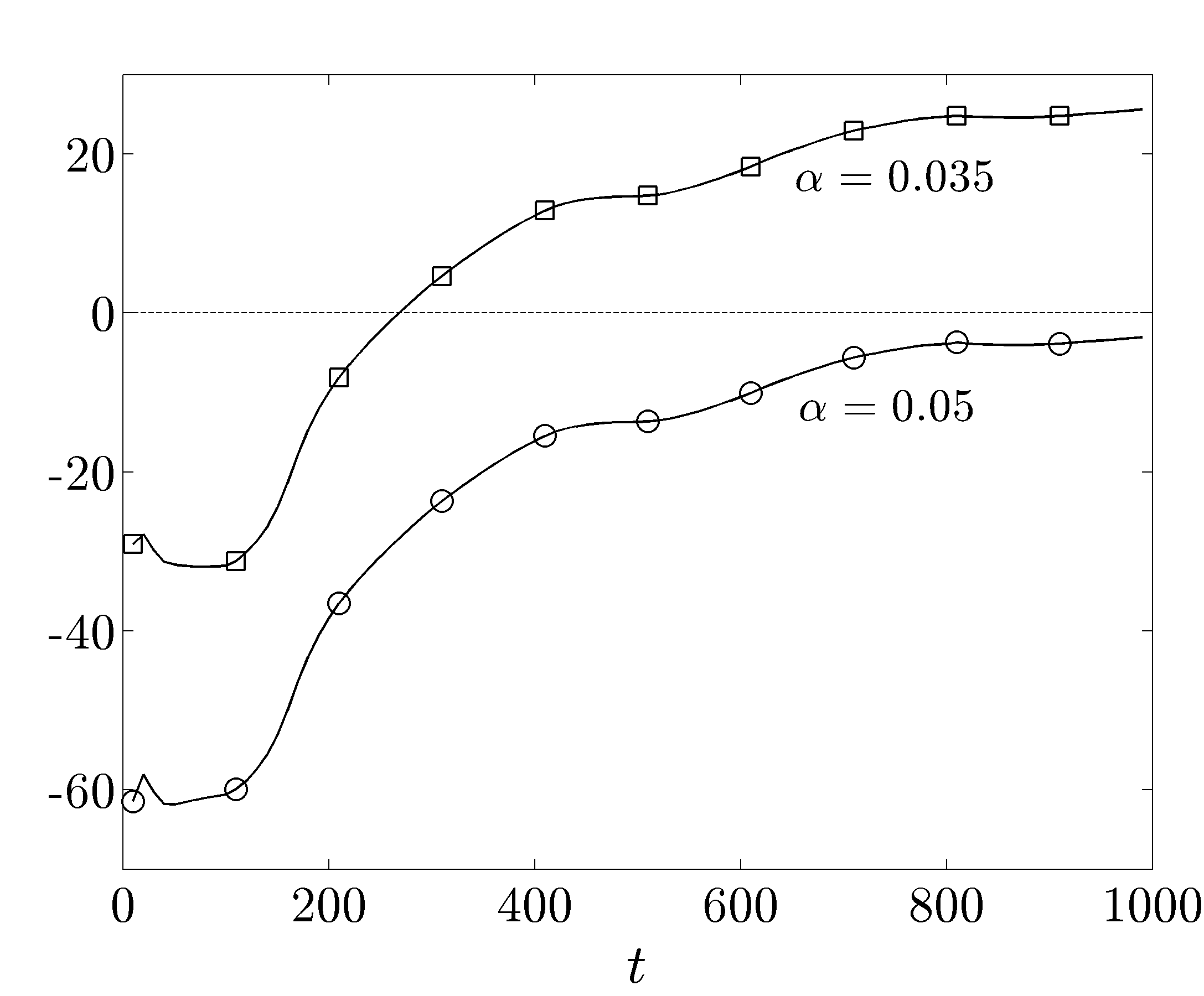}
                  \label{fig.Pnet-moderateIC-dtw}
               }
            \end{tabular}
         \end{center}
         \caption{(a) Energy of the velocity fluctuations, $E (t)$; (b) skin-friction drag coefficient, $C_f (t)$; (c) normalized required power, $\% \Pi_{req}$ (solid), normalized produced power, $\% \Pi_{prod}$ (dashed); and (d) normalized net power, $\% \Pi_{net}$, for the initial condition with moderate energy: DTWs with $\square$,~$(c = 5, \omega_{x} = 2, \alpha = 0.035)$; $\circ$,~$(c = 5, \omega_{x} = 2, \alpha = 0.05)$.}
         \label{fig.moderateIC-dtw}
      \end{figure}
		
		The fluctuations' kinetic energy and skin-friction drag coefficient for the DTWs are shown in figures~\ref{fig.FE-moderateIC-dtw} and \ref{fig.Cf-moderateIC-dtw}. Figure~\ref{fig.FE-moderateIC-dtw} shows that the DTWs with $(c = 5$,~$\omega_{x} = 2$,~$\alpha = \{0.035, 0.05\})$ significantly weaken intensity of the velocity fluctuations, thereby facilitating maintenance of the laminar flow. From figure~\ref{fig.Cf-moderateIC-dtw} we also see that the small transient growth of fluctuations' kinetic energy results in a small transient increase in the drag coefficients which eventually decay to their nominal values reported in table~\ref{table.nominal}. Even though these drag coefficients are larger than in the uncontrolled laminar flow, they are still approximately two times smaller than in the uncontrolled flow that becomes turbulent (cf.\ table~\ref{table.3D}).
		
		The normalized produced, required, and net powers for DTWs are shown in figures~\ref{fig.Preq-Psav-moderateIC-dtw} and \ref{fig.Pnet-moderateIC-dtw}. As can be seen from figure~\ref{fig.Preq-Psav-moderateIC-dtw}, the normalized produced power for the DTWs is positive by virtue of the fact that the uncontrolled flow becomes turbulent while the controlled flows stay laminar. Figure~\ref{fig.Pnet-moderateIC-dtw} shows that the DTW with $\alpha = 0.035$ (respectively, $\alpha = 0.05$) has a positive (respectively, negative) net power balance. The reason for this is twofold: first, as evident from figure~\ref{fig.Preq-Psav-moderateIC-dtw}, the DTW with larger $\alpha$ results in a smaller produced power since it induces a larger negative nominal bulk flux than the DTW with smaller $\alpha$; and second, the required power to maintain the DTW with larger $\alpha$ is bigger than in the DTW with smaller $\alpha$. Furthermore, at $t = 1000$, the DTW with $\alpha = 0.035$ has a larger net power than the UTW with $\alpha = 0.015$ ($\% \Pi_{net} = 25.63$ vs.\ $\% \Pi_{net} = 6.83$; cf.\ table~\ref{table.3D}). This is because the DTW with $\alpha = 0.035$, in contrast to the UTW with $\alpha = 0.015$, remains laminar and produces a much larger power than it requires.

In summary, the results of this section highlight an important trade-off that  needs to be taken into account when designing the traveling waves. Large amplitudes of properly designed downstream waves yield larger receptivity reduction which is desirable for controlling the onset of turbulence. However, this is accompanied by an increase in drag coefficient and required control power. Thus, to maximize net efficiency, it is advantageous to select the smallest possible amplitude of wall-actuation that can maintain the laminar flow.

   \subsection{Large initial energy}
    	\label{sec.large}

  		Section~\ref{sec.moderate} illustrates capability of properly designed DTWs to maintain the laminar flow in the presence of initial conditions that induce transition in the uncontrolled flow. In this section, we demonstrate that, as the energy of the initial perturbation increases, a DTW with larger amplitude is needed to prevent transition. Our results confirm the prediction made in Part~1 that maintaining a laminar flow with a larger DTW amplitude comes at the expense of introducing a negative net power balance.

\begin{figure}
        \begin{center}
            \begin{tabular}{cc}
               $E$
               &
               $C_f$
					\\[-0.15cm]
               \subfigure[]
               {
                  \includegraphics[width=0.47\columnwidth]
                  {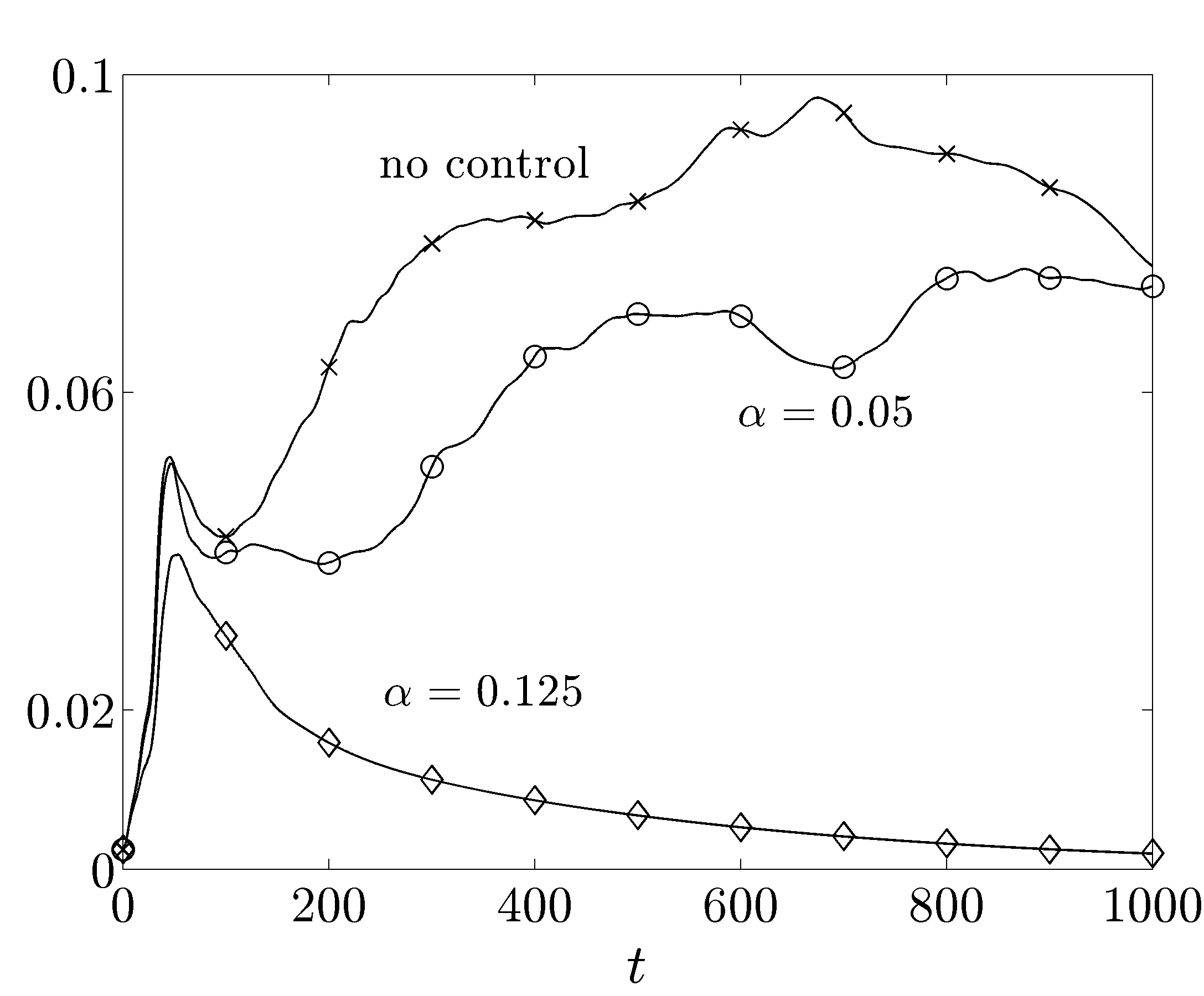}
                  \label{fig.FE-largeIC-dtw}
               }
               &
               \subfigure[]
               {
                  \includegraphics[width=0.47\columnwidth]
                  {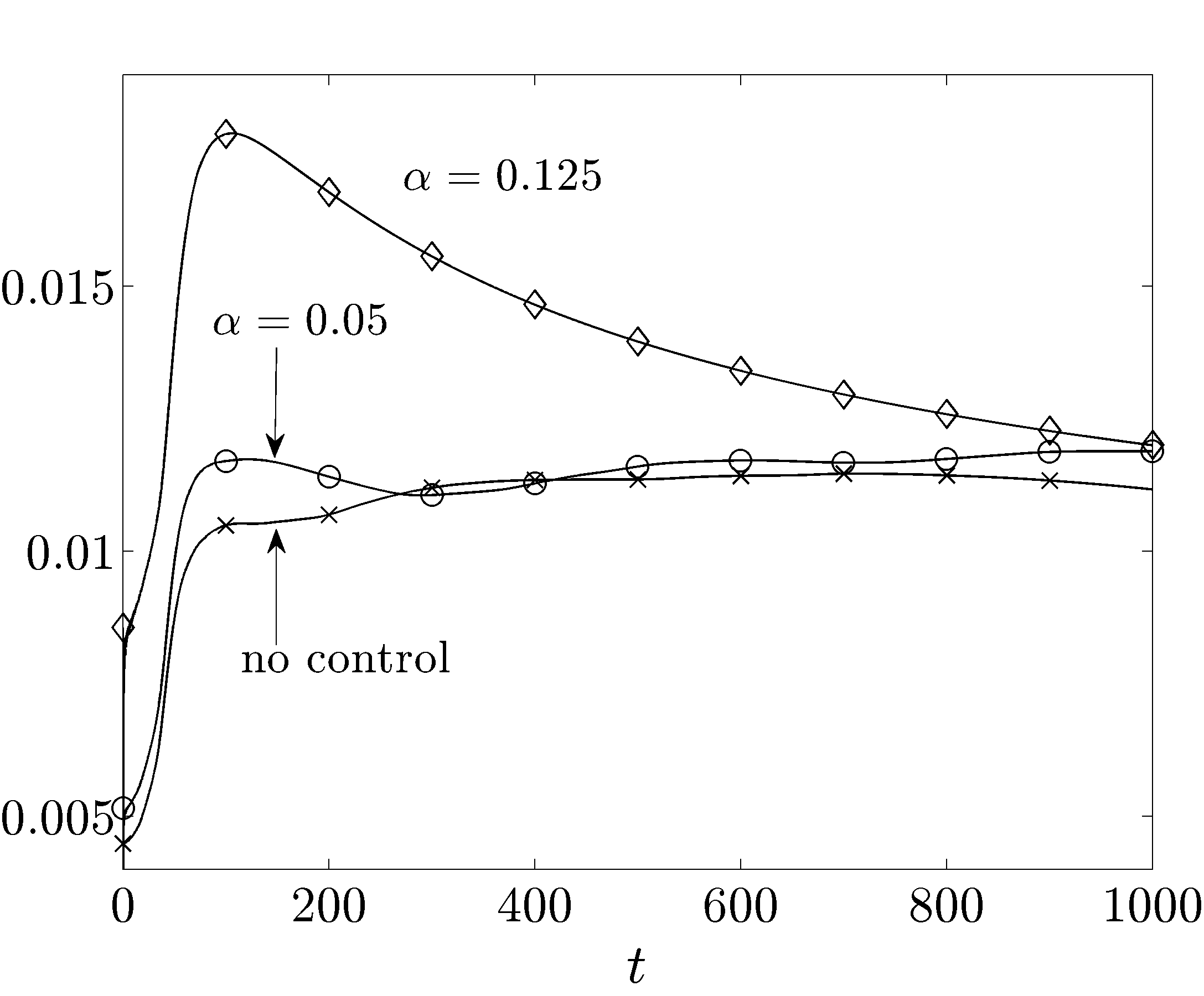}
                  \label{fig.Cf-largeIC-dtw}
               }
					\\
					$\% \Pi_{req}$, $\% \Pi_{prod}$
					&
					$\% \Pi_{net}$
					\\[-0.15cm]
					\subfigure[]
               {
                  \includegraphics[width=0.47\columnwidth]
                  {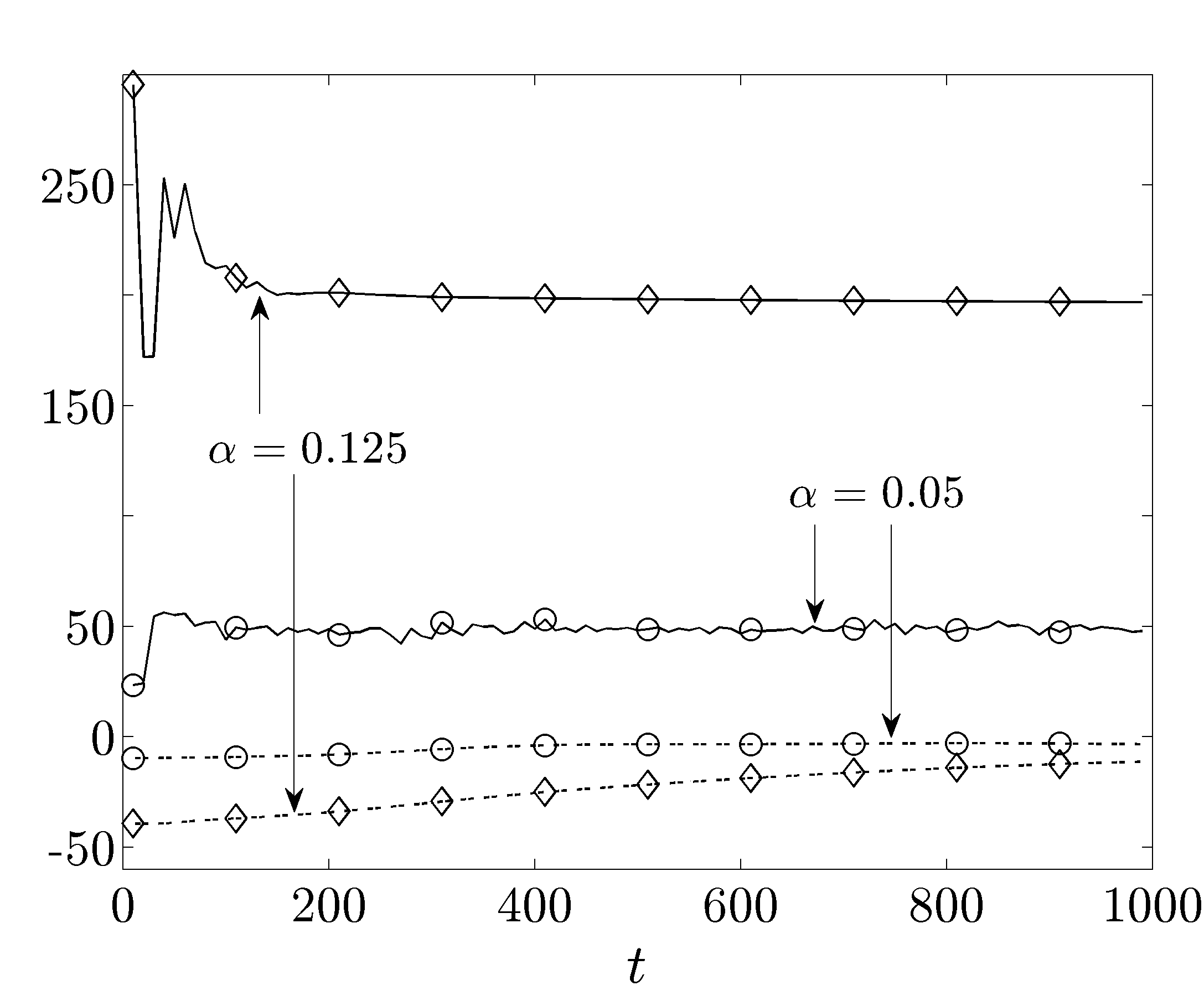}
                  \label{fig.Preq-Psav-largeIC-dtw}
               }
					&
					\subfigure[]
               {
                  \includegraphics[width=0.47\columnwidth]
                  {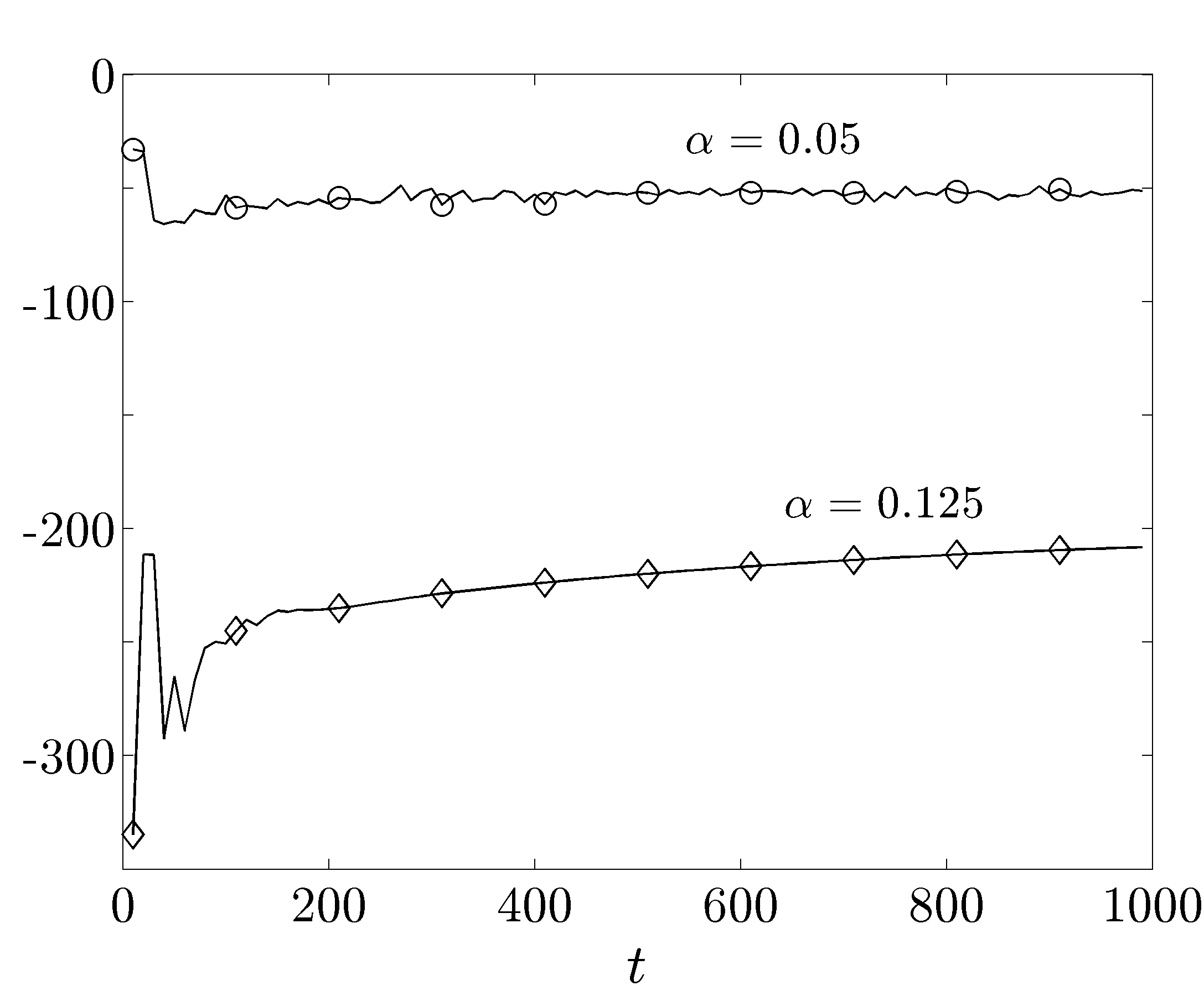}
                  \label{fig.Pnet-largeIC-dtw}
               }
            \end{tabular}
         \end{center}
         \caption{(a) Energy of the velocity fluctuations, $E (t)$; (b) skin-friction drag coefficient, $C_f (t)$; (c) normalized required power, $\% \Pi_{req}$ (solid), normalized produced power, $\% \Pi_{prod}$ (dashed); and (d) normalized net power, $\% \Pi_{net}$, for the initial condition with large energy: $\times$,~uncontrolled; DTWs with $\circ$,~$(c = 5, \omega_{x} = 2, \alpha = 0.05)$; and $\lozenge$,~$(c = 5$, $\omega_{x} = 2$, $\alpha = 0.125)$.}
         \label{fig.largeIC-dtw}
      \end{figure}

        Simulations in this section are done for the initial condition with large kinetic energy, $E (0) = 2.5 \times 10^{-3}$. The time evolution of the fluctuations' energy for a pair of DTWs with $(c = 5$, $\omega_{x} = 2$, $\alpha = \{ 0.05, 0.125 \})$ is shown in figure~\ref{fig.FE-largeIC-dtw}. The uncontrolled flow becomes turbulent and exhibits similar trends in the evolution of $E (t)$ as the corresponding flow initiated with moderate energy perturbations (cf.\ figures~\ref{fig.FE-moderateIC-utw} and~\ref{fig.FE-largeIC-dtw}). On the other hand, figure~\ref{fig.FE-largeIC-dtw} shows that the DTW with $\alpha = 0.05$ is not capable of maintaining the laminar flow; in comparison, the same set of control parameters prevented transition for the perturbations of moderate initial energy (cf.\ figures~\ref{fig.FE-moderateIC-dtw} and~\ref{fig.FE-largeIC-dtw}). Conversely, the DTW with $\alpha = 0.125$ remains laminar even though $E (t)$ transiently reaches about half the energy of the turbulent uncontrolled flow. Therefore, the DTWs with frequency and speed selected in Part~1 and sufficiently large amplitudes are capable of maintaining the laminar flow even in the presence of large initial perturbations.

        Figure~\ref{fig.Cf-largeIC-dtw} shows the skin-friction drag coefficients for the flows considered in figure~\ref{fig.FE-largeIC-dtw}. For the DTW with $\alpha = 0.05$ the steady-state value of $C_f$ is given by $C_f = 11.9 \times 10^{-3}$, which is a slightly larger value than in the turbulent uncontrolled flow, $C_f = 11.2 \times 10^{-3}$ (cf.\ table~\ref{table.3D}). We note that the drag coefficient of the DTW that stays laminar initially reaches values that are about $50 \, \%$ larger than in the uncontrolled flow; after this initial increase, $C_f (t)$ then gradually decays to the value predicted using the base flow analysis, $C_f = 8.6 \times 10^{-3}$  (cf.\ table~\ref{table.nominal}). Figures~\ref{fig.Preq-Psav-largeIC-dtw} and~\ref{fig.Pnet-largeIC-dtw} show the normalized required, produced, and net powers for the initial condition with large kinetic energy. As evident from figure~\ref{fig.Pnet-largeIC-dtw}, the net power balance is negative for all considered flows. The DTW with $\alpha = 0.05$ becomes turbulent, and it has a larger drag coefficient than the uncontrolled flow which consequently leads to negative produced and net powers. Moreover, even though the DTW with $\alpha = 0.125$ can sustain laminar flow, its net power balance is very poor. There are two main reasons for the lack of efficiency of this control strategy: first, its nominal drag coefficient is significantly larger than in a DTW with smaller amplitudes which consequently yields very small produced power (at larger times not shown in figure~\ref{fig.Preq-Psav-largeIC-dtw}); and second, a prohibitively large power is required to maintain this large amplitude DTW.


        The results of this section show that preventing transition by DTWs in the presence of large initial conditions comes at the expense of large negative net power balance. We also highlight that in the presence of large initial perturbations (or, equivalently, at large Reynolds numbers), transition to turbulence may be inevitable. Furthermore, the results of~\S~\ref{sec.moderate} show that the UTWs may reduce the turbulent skin-friction drag and achieve positive net efficiency. The approach used in Part~1 considers dynamics of fluctuations around laminar flows and, thus, it cannot be used for explaining the positive efficiency of the UTWs that become turbulent.

	\subsection{Energy amplification mechanisms}
   \label{sec.mech-energy}

  		\begin{figure}
      	\begin{center}
            \begin{tabular}{cc}
					{\sc downstream:} $P_E$, $D_E$
					&
               {\sc upstream:} $P_E$, $D_E$
               \\[-0.2cm]
					\subfigure[]
               {
                  \includegraphics[width=0.47\columnwidth]
                  {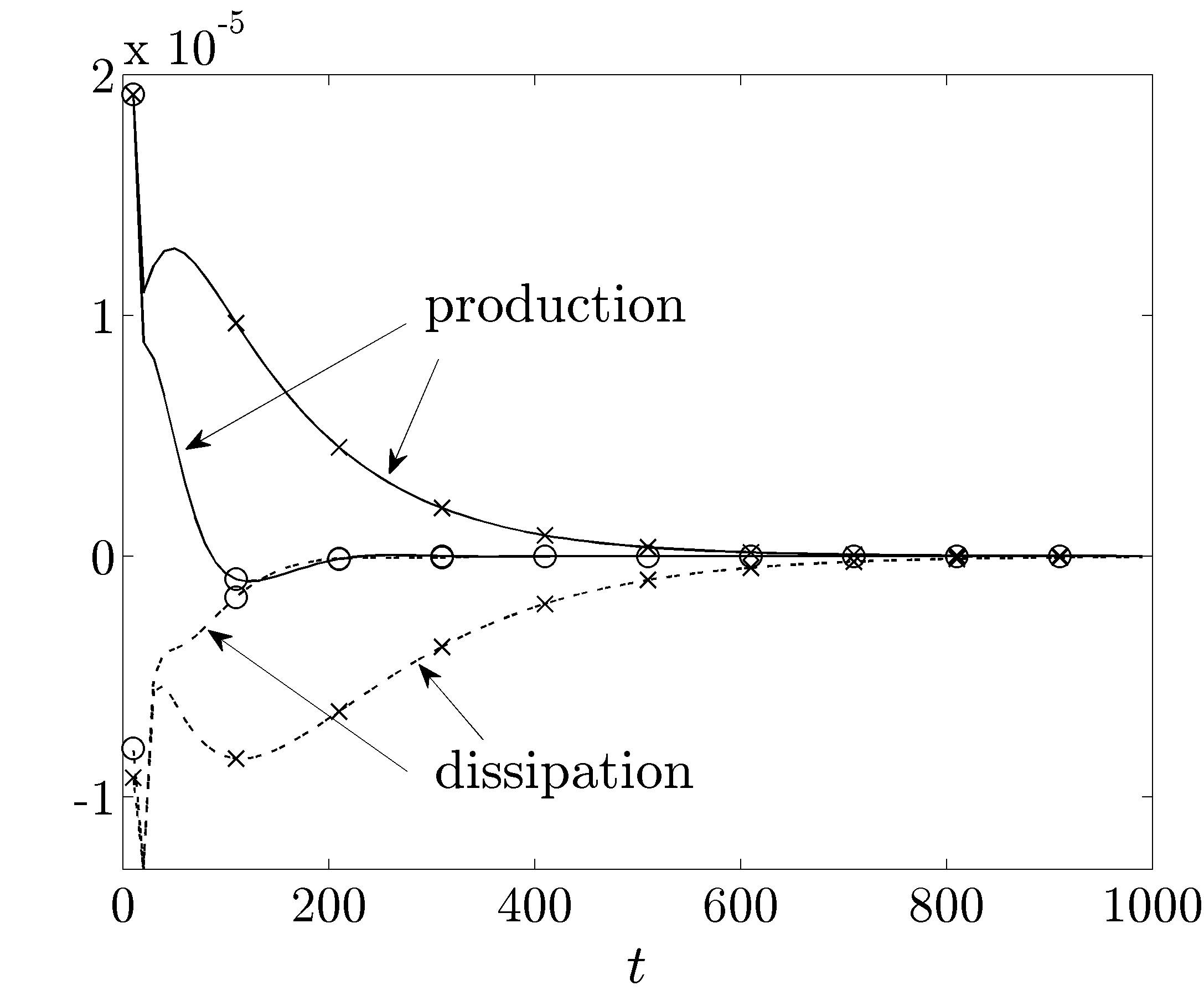}
                  \label{fig.P-D-DTW-small}
               }
					&
               \subfigure[]
               {
                  \includegraphics[width=0.47\columnwidth]
                  {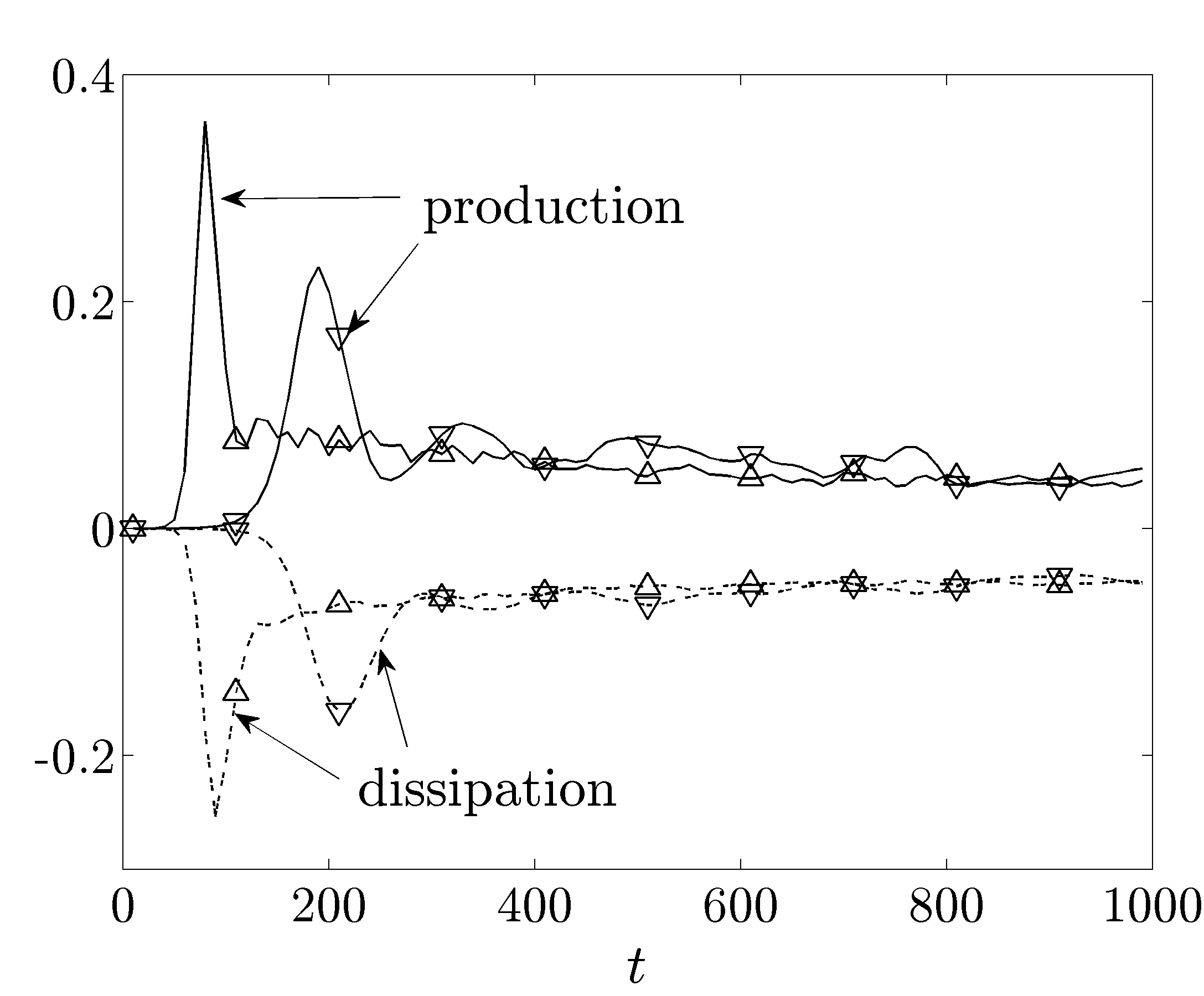}
                  \label{fig.P-D-UTW-small}
               }
				\end{tabular}
				\begin{tabular}{c}
					{\sc upstream:} $P_E + D_E$
					\\[-0.2cm]
               \subfigure[]
               {
                  \includegraphics[width=0.47\columnwidth]
                  {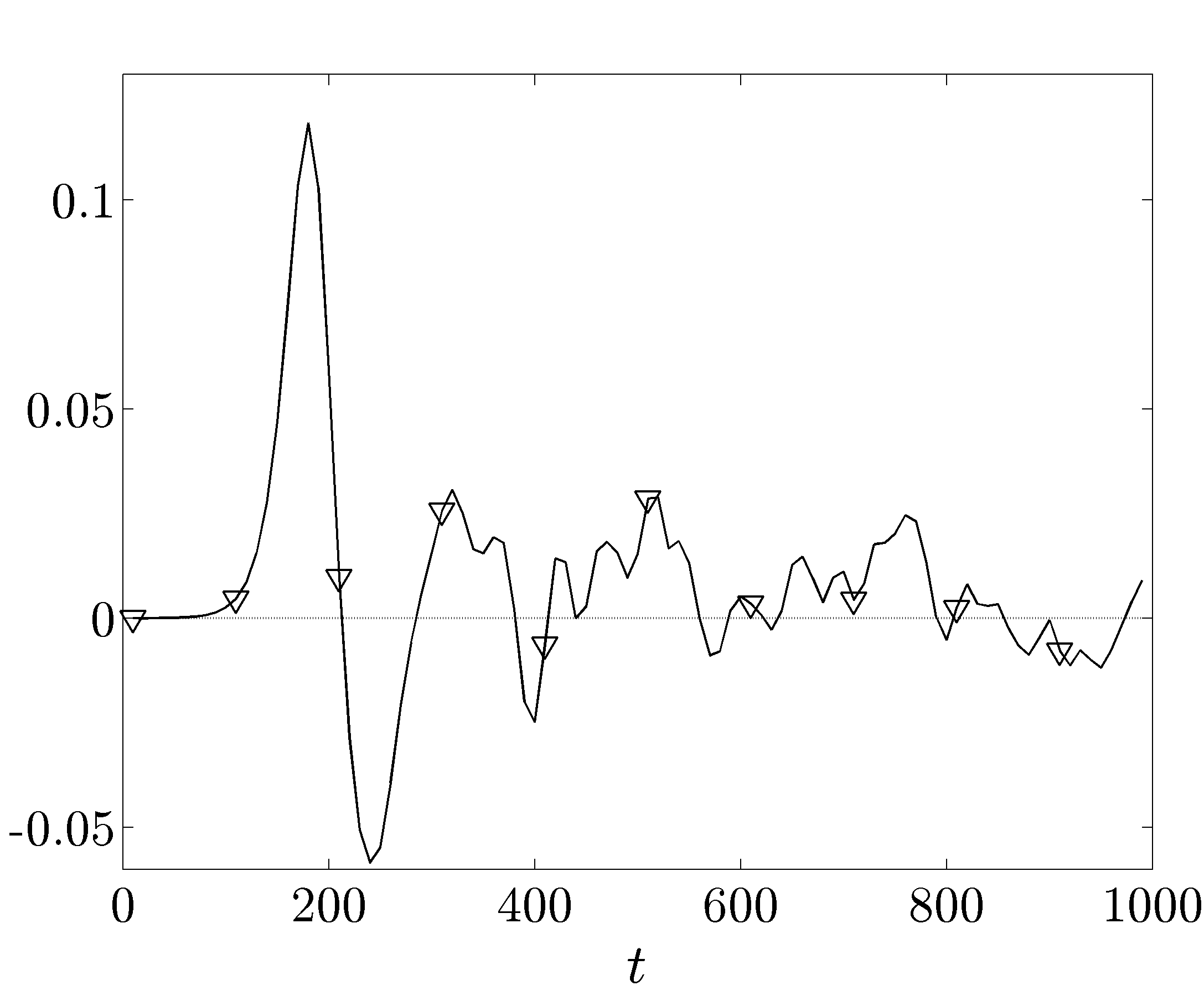}
                  \label{fig.Et_UTW_small}
               }
            \end{tabular}
         \end{center}
         \caption{(a) and (b) Production, $P_E (t)$ (solid), and dissipation, $D_E (t)$ (dashed), of kinetic energy in Poiseuille flow with $R_{\tau} = 63.25$ for the initial condition with small energy:
    (a) $\times$, uncontrolled; $\circ$, DTW with~$(c = 5$, $\omega_{x} = 2$, $\alpha = 0.05)$; and
    (b) $\triangledown$, UTW with~$(c = -2$, $\omega_{x} = 0.5$, $\alpha = 0.05)$; $\vartriangle$, UTW with~$\alpha = 0.125$.
    (c) $P_E (t) + D_E (t)$ for the UTW with~$(c = -2$, $\omega_{x} = 0.5$, $\alpha = 0.05)$.}
         \label{fig.P-D-small}
      \end{figure}

		The Reynolds-Orr equation can be used to quantify the evolution of fluctuations' kinetic energy around a given base flow~\citep{schhen01}. In this section, we use this equation to examine mechanisms that contribute to production and dissipation of kinetic energy in flows subject to traveling waves. For base velocity, $\bu_{b} = (U(x, y, t), V(x, y, t), 0)$, the evolution of the energy of velocity fluctuations, $E (t)$, is determined by
        \be
        \ba{rcl}
        \dfrac{1}{2} \, \dfrac{\mrd E}{\mrd t}
        \!\! & = & \!\!
        P_E (t)
        \, + \,
        D_E (t),
        \\[0.25cm]
        P_E (t)
        \!\! & = & \!\!
        -\dfrac{1}{\Omega} \,
        \ds{\int_{\Omega}{
        \left(
        u v U_{y}
        \, + \,
        u v V_{x}
        \, + \,
        v^{2} V_{y}
        \, + \,
        u^{2} U_{x}
        \right)
        \mrd \Omega}},
        \\[0.4cm]
        D_E (t)
        \!\! & = & \!\!
        \dfrac{1}{R_c \Omega} \,
        \ds{\int_{\Omega}{\bv \cdot \Delta \bv \, \mrd \Omega}}.
        \ea
        \label{eq.evol-energy}
        \ee
Here, $P_E$ represents the production of kinetic energy and is associated with the work of the Reynolds stresses on the base shear, whereas $D_E$ accounts for viscous dissipation.

		We confine our attention to the simulations for initial conditions with small energy. This situation is convenient for explaining why the DTWs exhibit improved transient behavior compared to the laminar uncontrolled flow while the UTWs promote transition to turbulence. Figure~\ref{fig.P-D-DTW-small} shows production and dissipation terms for the uncontrolled flow and for the flow subject to a DTW with ($c = 5$, $\omega_{x} = 2$, $\alpha = 0.05$). For the uncontrolled flow, $P_{E}$ is always positive, $D_E$ is always negative, and they both decay to zero at large times. On the contrary, the production term for the DTW becomes negative for $80 \lesssim t \lesssim 220$. We see that, at early times, $P_E$ and $D_E$ for the DTW follow their uncontrolled flow counterparts. However, after this initial period, they decay more rapidly to zero. These results confirm the prediction of Part~1 that the DTWs reduce the production of kinetic energy. In contrast, figure~\ref{fig.P-D-UTW-small} shows that the UTWs with ($c = -2$, $\omega_{x} = 0.5$, $\alpha = \{0.05, 0.125\}$) increase both $P_E$ and $D_E$ by about four orders of magnitude compared to the values reported in figure~\ref{fig.P-D-DTW-small}. This verifies the theoretical prediction of Part~1 that the UTWs increase the production of kinetic energy. Moreover, production dominates dissipation transiently, thereby inducing large growth of kinetic energy observed in figure~\ref{fig.FE-smallIC-utw}. For the UTW with $\alpha = 0.05$, this is further illustrated in figure~\ref{fig.Et_UTW_small} by showing that $P_E$ accumulates more energy than $D_E$ dissipates (i.e., the area under the curve in figure~\ref{fig.Et_UTW_small} is positive). We also note that, in the above simulations, the work of Reynolds stress $u v$ on the base shear $U_{y}$ dominates the other energy production terms. Furthermore, our results show that the wall-normal diffusion of $u$ is responsible for the largest viscous energy dissipation.

	\subsection{Flow visualization}
		\label{sec.flow-structures}
		
		In sections~\ref{sec.small} -~\ref{sec.large}, transition was identified by examining fluctuations' kinetic energy and skin-friction drag coefficients. Large levels of sustained kinetic energy and substantial increase in drag coefficients (compared to base flows) were used as indicators of transition. Here, we use flow visualization to identify coherent structures in both the uncontrolled and controlled flows.
	
\begin{figure}
		   \begin{center}
		      \begin{tabular}{ccc}
					{\sc uncontrolled:} & {\sc upstream:} & {\sc downstream:}
					\\
					$t = 50$ & $t = 50$ & $t = 50$
					\\[-0.2cm]
		         \subfigure[]
		         {
		            \includegraphics[width=0.315\textwidth]
		            {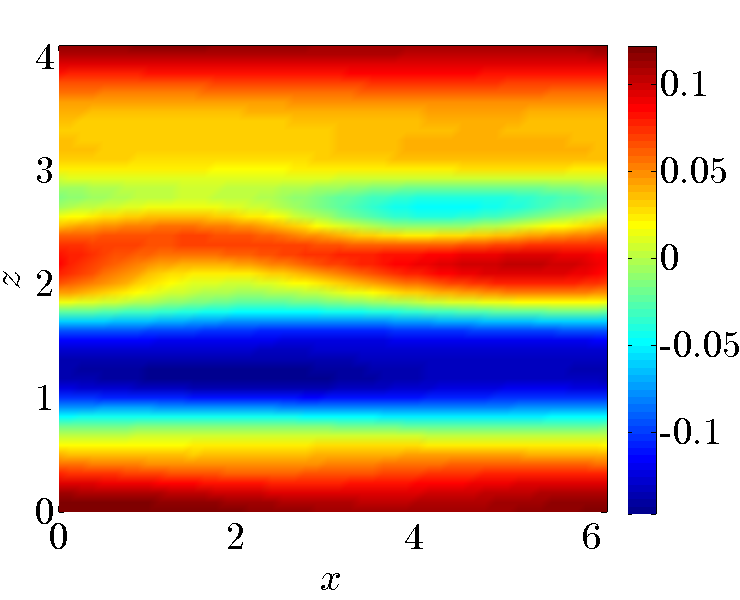}
		            \label{fig.unc_u_y45_t50}
		         }
		         &
		         \subfigure[]
		         {
		            \includegraphics[width=0.315\textwidth]
		            {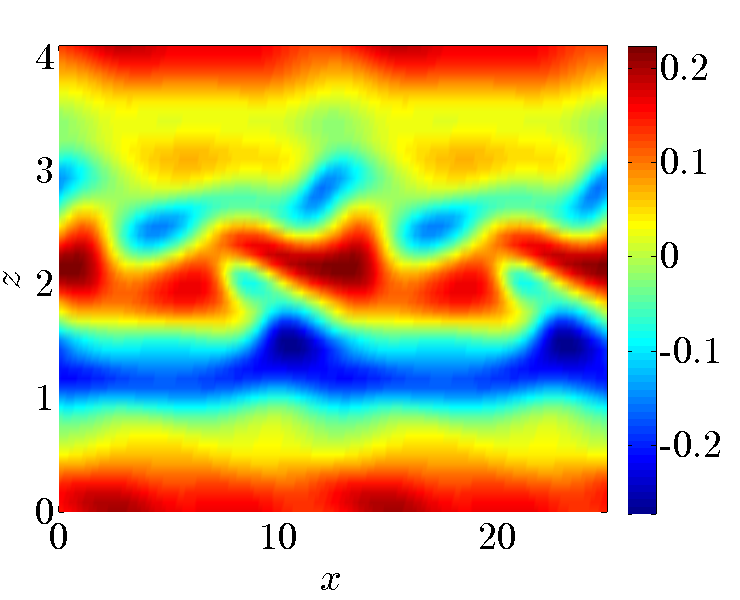}
		            \label{fig.utw_u_y45_t50}
		         }
					&
					\subfigure[]
		         {
		            \includegraphics[width=0.315\textwidth]
		            {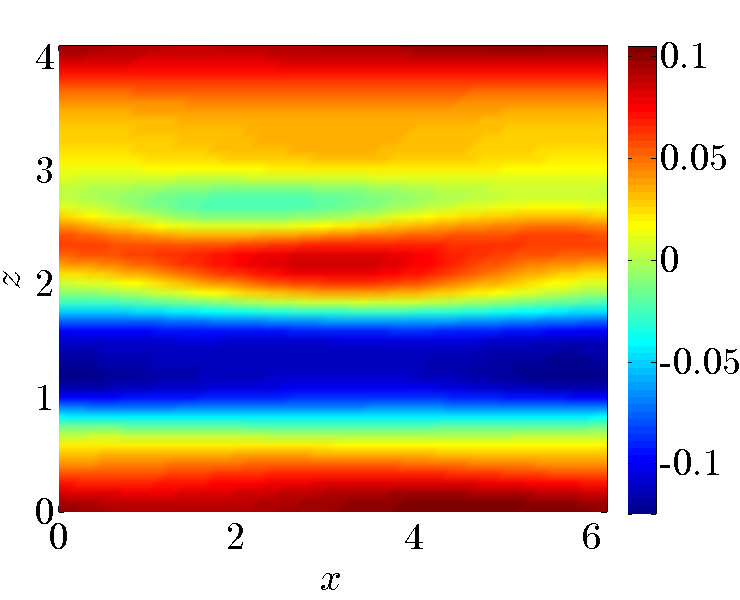}
		            \label{fig.dtw_u_y45_t50}
		         }
					\\
					$t = 120$ & $t = 120$ & $t = 120$
					\\[-0.2cm]
		         \subfigure[]
		         {
		            \includegraphics[width=0.315\textwidth]
		            {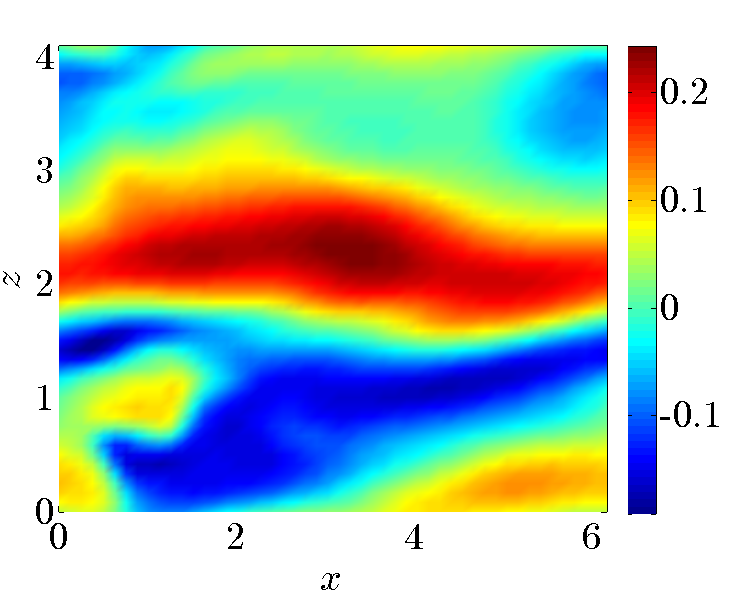}
		            \label{fig.unc_u_y45_t120}
		         }
		         &
		         \subfigure[]
		         {
		            \includegraphics[width=0.315\textwidth]
		            {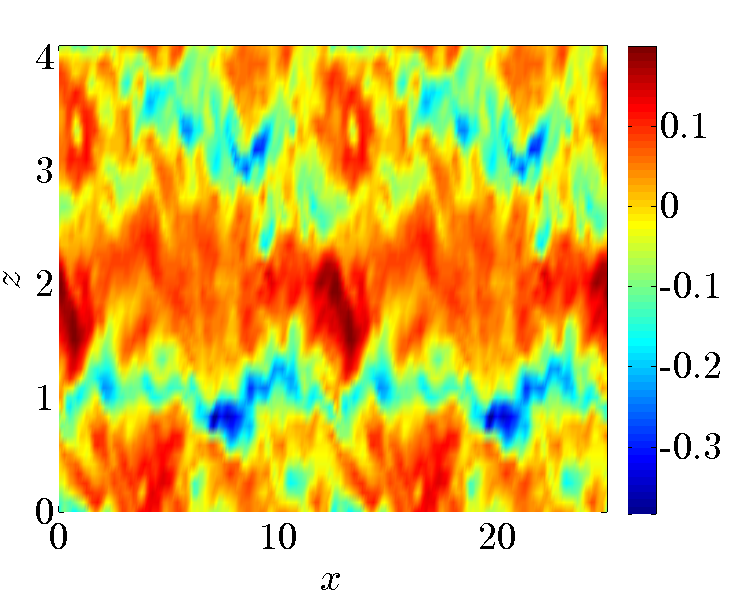}
		            \label{fig.utw_u_y45_t120}
		         }
					&
					\subfigure[]
		         {
		            \includegraphics[width=0.315\textwidth]
		            {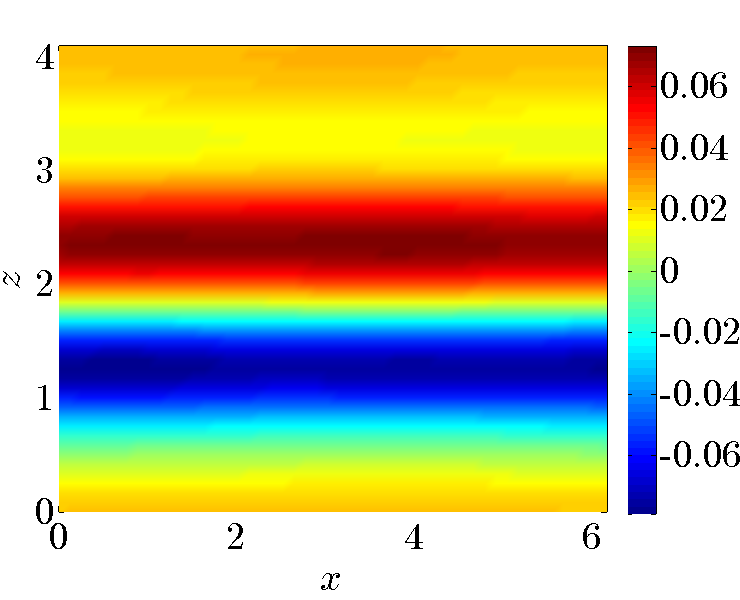}
		            \label{fig.dtw_u_y45_t120}
		         }
		      \end{tabular}
		   \end{center}
		   \caption{Streamwise velocity fluctuations, $u(x,z)$, at $y = -0.5557$ ($y^+ = 28.11$), (a) - (c) $t = 50$, and (d) - (f) $t = 120$ for initial condition with moderate energy: uncontrolled flow; UTW with~$(c = -2$, $\omega_{x} = 0.5$, $\alpha = 0.05)$; and DTW with~$(c = 5$, $\omega_{x} = 2$, $\alpha = 0.05)$.}
		   \label{fig.moderateIC_u_y45}
		\end{figure}

		The onset of turbulence in a bypass transition is usually characterized by formation of streamwise streaks and their subsequent break-down. For the initial condition with moderate energy, figure~\ref{fig.moderateIC_u_y45} shows the streamwise velocity fluctuations at $y = -0.5557$ ($y^+ = 28.11$ in wall units) for the uncontrolled flow and flows subject to a UTW with ($c = -2$, $\omega_x = 0.5$, $\alpha = 0.05$) and a DTW with ($c = 5$, $\omega_x = 2$, $\alpha = 0.05$). Clearly, the initial perturbations evolve into streamwise streaks in all three flows (cf. figures~\ref{fig.unc_u_y45_t50} -~\ref{fig.dtw_u_y45_t50} for $t = 50$). At $t = 120$, the growth of velocity fluctuations results in a break-down of the streaks both in the flow with no control and in the flow subject to UTWs (cf.\ figures~\ref{fig.unc_u_y45_t120} and~\ref{fig.utw_u_y45_t120}). We see that the streaks evolve into complex flow patterns much faster in the latter case. For the UTWs, the streak distortion occurs as early as $t = 50$ and a broad range of spatial scales is observed at $t = 120$. On the contrary, figure~\ref{fig.dtw_u_y45_t120} shows that, at $t = 120$, the DTWs have reduced the magnitude of velocity fluctuations to about half the value at $t = 50$, thereby weakening intensity of the streaks and maintaining the laminar flow.
	
\section{Relaminarization by downstream waves}
	\label{sec.relaminarization}

Thus far we have shown that properly designed DTWs represent an effective means for controlling the onset of turbulence. In this section, we demonstrate that the DTWs designed in Part 1 can also relaminarize fully developed turbulent flows. Since the lifetime of turbulence depends on the Reynolds number~\citep{bro89,gro00,hofwesscheck06,borschtag10}, we examine turbulent flows with $R_c = 2000$ (i.e., $R_\tau \approx 63.25$) and $R_c = 4300$ (i.e., $R_\tau \approx 92.80$).
		
		\begin{figure}
         \begin{center}
                \begin{tabular}{c}
                  \includegraphics[width=0.47\columnwidth]
                  {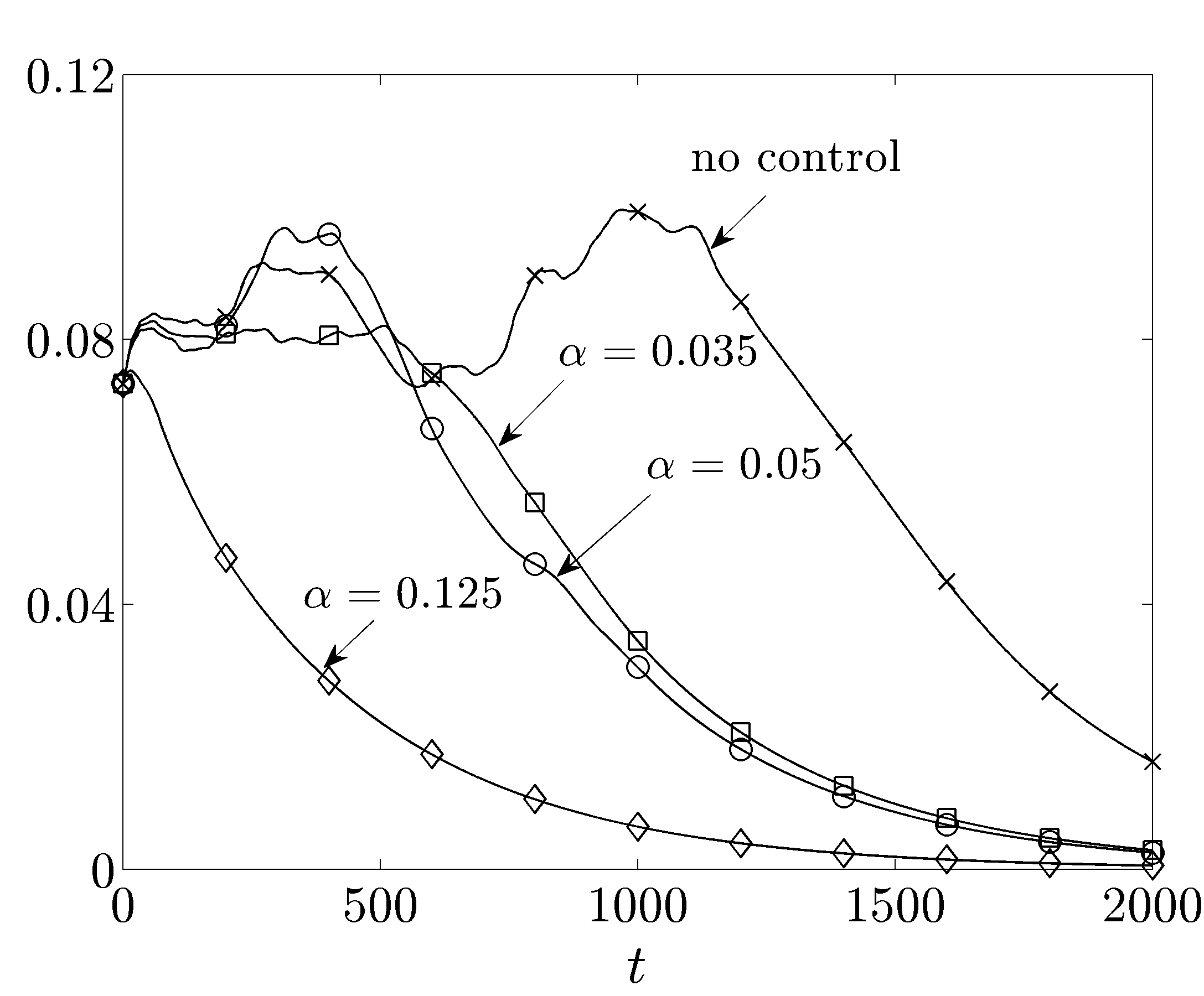}
                \end{tabular}
         \end{center}
          \caption{Energy of velocity fluctuations around base flows of \S~\ref{sec.goveq}. Simulations are initiated by a fully developed turbulent flow with $R_c = 2000$:~$\times$, uncontrolled; DTWs with $\square$,~$(c = 5, \omega_{x} = 2, \alpha = 0.035)$; $\circ$,~$(c = 5, \omega_{x} = 2, \alpha = 0.05)$; and $\lozenge$,~$(c = 5$, $\omega_{x} = 2$, $\alpha = 0.125)$.}
          \label{fig.energy_dtw_Re2000}
      \end{figure}

		The numerical scheme described in \S~\ref{sec.numerical_method} is used to simulate the turbulent flows. For $R_c = 4300$, the number of grid points in the streamwise, wall-normal, and spanwise directions is increased to $80 \times 97 \times 80$. The velocity field is initialized with a fully developed turbulent flow obtained in the absence of control. The surface blowing and suction that generates DTWs is then introduced (at $t = 0$), and the kinetic energy and drag coefficients are computed at each time step.
		
		The fluctuations' kinetic energy for the uncontrolled flow and for the flows subject to DTWs with $(c = 5$, $\omega_{x} = 2$, $\alpha = \{ 0.035, \, 0.05, \, 0.125 \})$ at $R_c = 2000$ are shown in figure~\ref{fig.energy_dtw_Re2000}. The energy of velocity fluctuations around base flows $\bu_b$ of \S~\ref{sec.goveq} (parabola for flow with no control; traveling waves for flow with control) are shown; relaminarization occurs when the energy of velocity fluctuations converges to zero. Clearly, large levels of fluctuations in the flow with no control are maintained up until $t \approx 1100$. After this time instant, however, velocity fluctuations exhibit gradual decay. On the other hand, fluctuations in flows subject to DTWs start decaying much earlier, thereby indicating that the lifetime of turbulence is reduced by surface blowing and suction. Relative to the uncontrolled flow, the fluctuations' kinetic energy for the DTWs considered here converge much faster to zero. We also see that the rate of decay increases as the wave amplitude gets larger.
		
		\begin{figure}
         \begin{center}
            \begin{tabular}{cc}
					$E$
               &
               $C_f$
					\\[-0.15cm]
               \subfigure[]
               {
                  \includegraphics[width=0.47\columnwidth]
                  {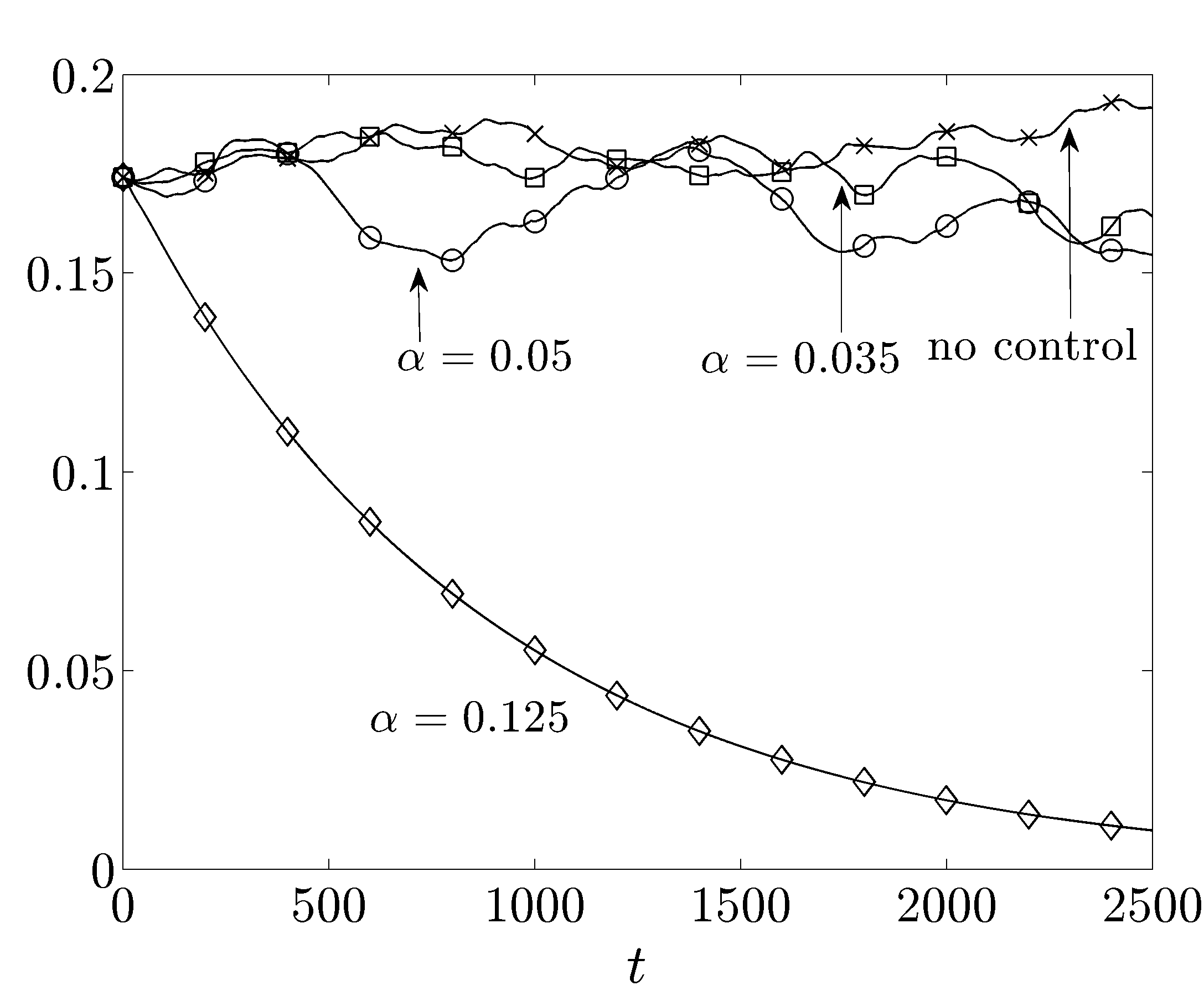}
                  \label{fig.energy_dtw_Re4300}
               }
               &
               \subfigure[]
               {
                  \includegraphics[width=0.47\columnwidth]
                  {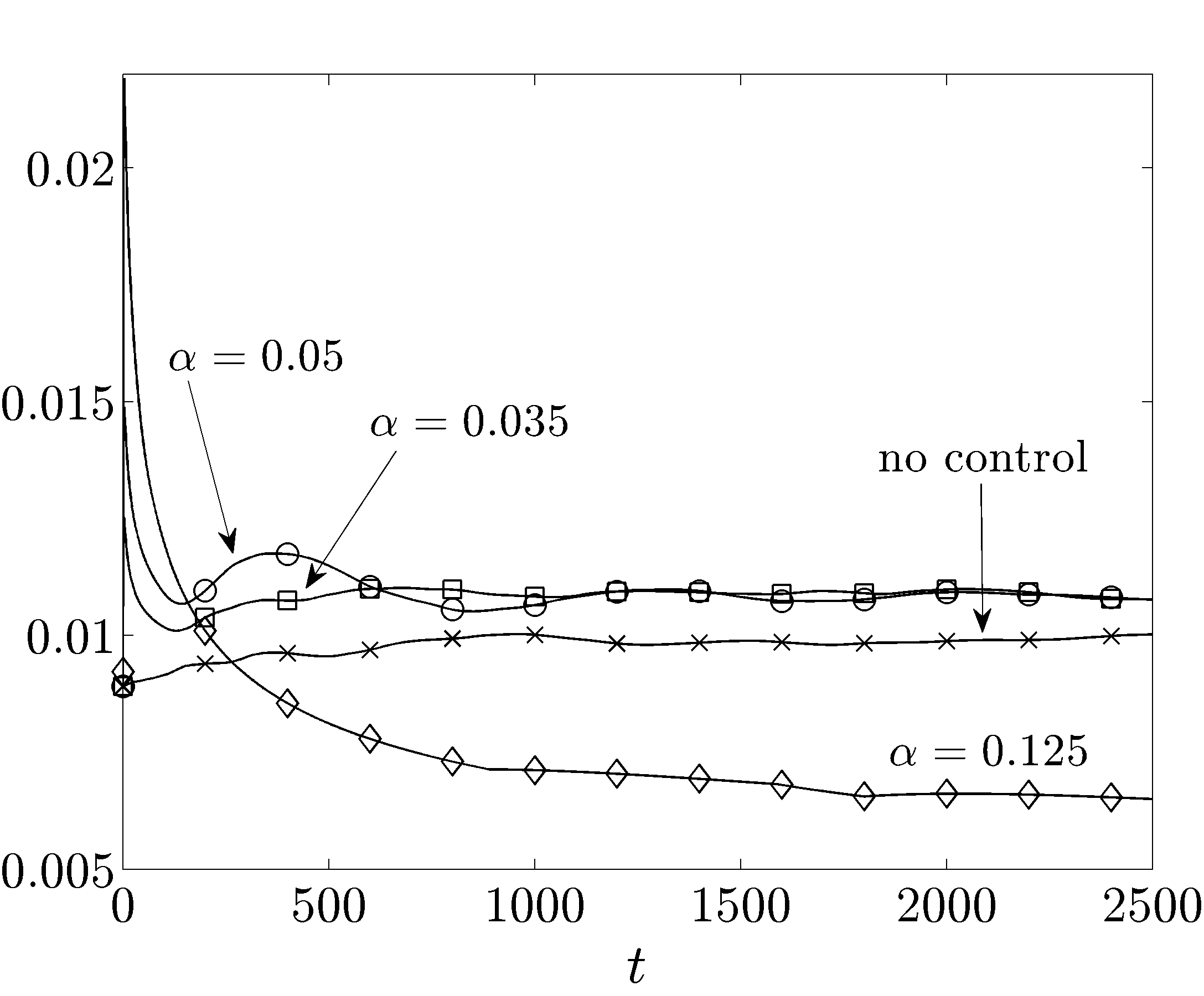}
                  \label{fig.Cf_dtw_Re4300}
               }
            \end{tabular}
         \end{center}
         \caption{(a) Energy of velocity fluctuations around base flows of \S~\ref{sec.goveq}, $E(t)$; and (b) skin-friction drag coefficient, $C_f(t)$. Simulations are initiated by a fully developed turbulent flow with $R_c = 4300$: $\times$, uncontrolled; DTWs with $\square$,~$(c = 5, \omega_{x} = 2, \alpha = 0.035)$; $\circ$,~$(c = 5, \omega_{x} = 2, \alpha = 0.05)$; and $\lozenge$,~$(c = 5$, $\omega_{x} = 2$, $\alpha = 0.125)$. }
         \label{fig.dtw_Re4300}
      \end{figure}

		We next consider a turbulent flow with $R_c = 4300$. The energy of velocity fluctuations around base flows $\bu_b$ of \S~\ref{sec.goveq} is shown in figure~\ref{fig.energy_dtw_Re4300}. In both the uncontrolled flow and the flows subject to the DTWs with $(c = 5$, $\omega_{x} = 2$, $\alpha = \{ 0.035, 0.05\})$ the energy oscillates around large values that identify turbulent flow. This indicates that the DTWs with smaller amplitudes cannot eliminate turbulence. On the contrary, the DTW with $\alpha = 0.125$ reduces the energy of velocity fluctuations, thereby relaminarizing the flow. Figure~\ref{fig.Cf_dtw_Re4300} shows that the skin-friction drag coefficient for the uncontrolled flow and for the DTWs with smaller amplitudes is approximately constant throughout the simulation. On the other hand, owing to relaminarization, the drag coefficient for the DTW with $\alpha = 0.125$ is smaller than that of the uncontrolled turbulent flow. However, relaminarization comes at the expense of poor net efficiency. This is because of the large required power (i.e., high cost of control), which reduces the appeal of using DTWs for control of turbulent flows.
	
\begin{figure}
         \begin{center}
               {
                  \includegraphics[width=0.8\columnwidth]
                  {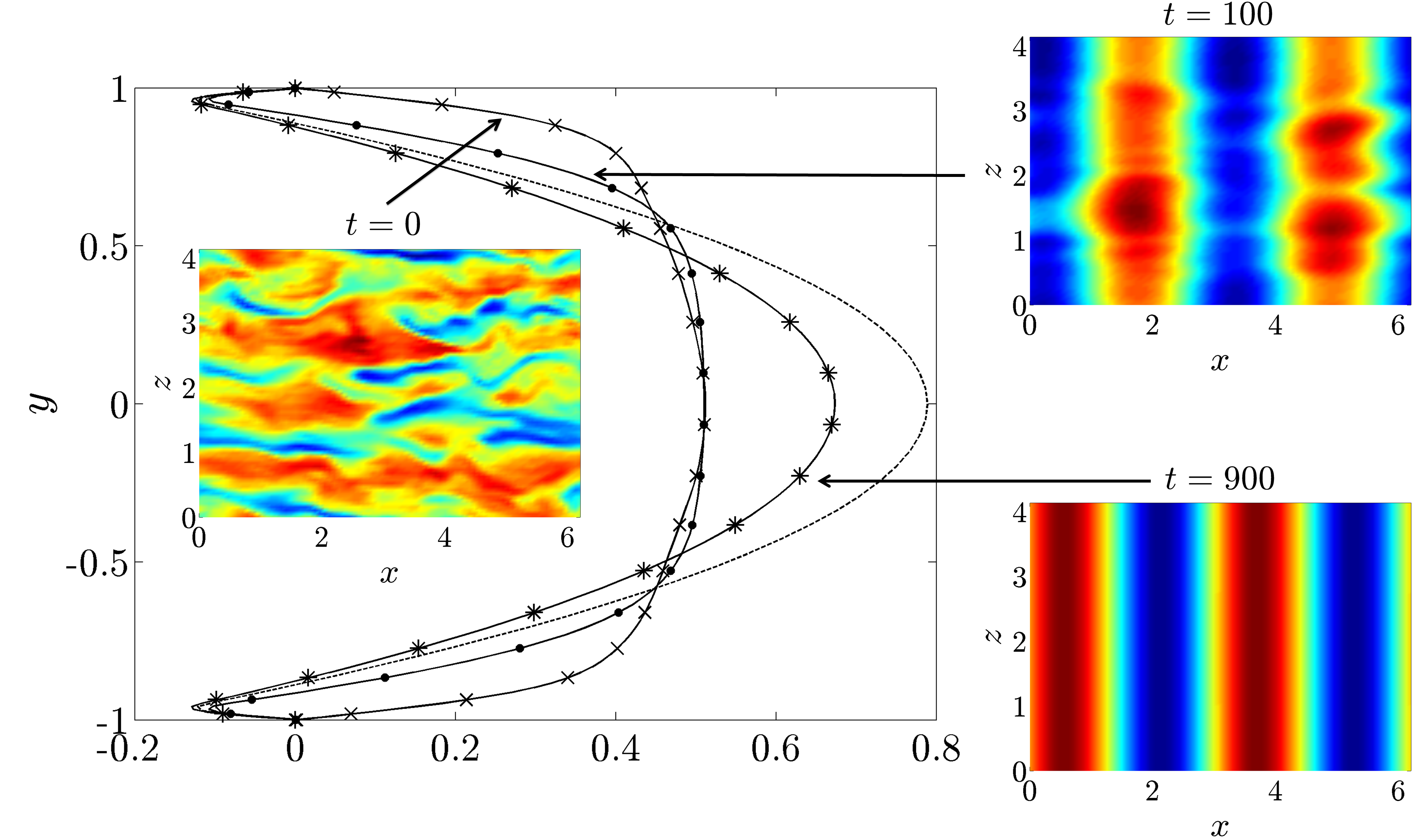}
               }
         \end{center}
         \caption{Mean velocity, $\overline{U}(y)$, and streamwise velocity (color plots) at $y = -0.7518$. Simulations are initiated by a fully developed turbulent flow with $R_c = 4300$: $\times$, $t = 0$; $\bullet$, $t = 100$; *, $t = 900$. The dashed line identifies the laminar mean velocity induced by the DTW with $(c = 5$, $\omega_{x} = 2$, $\alpha = 0.125)$.}
         \label{fig.umean_u_dtw_Re4300}
      \end{figure}

		Figure~\ref{fig.umean_u_dtw_Re4300} shows the mean velocity, $\overline{U}(y)$, at three time instants in the flow subject to the DTW with $(c = 5$, $\omega_{x} = 2$, $\alpha = 0.125)$. The instantaneous values of streamwise velocity in the ($x,z$)-plane at $y = -0.7518$ are also shown. As time advances, the initial turbulent mean velocity at $R_c = 4300$ moves towards the laminar mean velocity induced by the surface blowing and suction (dashed line). The color plots illustrate how the initial turbulent flow evolves into the DTW laminar base flow. We conjecture that receptivity reduction is important not only for controlling the onset of turbulence but also for relaminarization of fully developed flows. As outlined in Part~1, explaining the effect of traveling waves on turbulent flows requires additional control-oriented modeling and further scrutiny.

\section{Concluding remarks}
   \label{sec.conclusion}

This work, along with a companion paper~\citep{moajov10}, represents a continuation of recent efforts~\citep*{jovmoayouCTR06,jovPOF08} to develop a {\em model-based\/} paradigm for design of sensorless flow control strategies in wall-bounded shear flows. Direct numerical simulations are used to complement and verify theoretical predictions of~\cite{moajov10}, where receptivity analysis of the linearized NS equations was used to design small amplitude traveling waves. We have shown that perturbation analysis (in the wave amplitude) represents a powerful simulation-free method for predicting full-scale phenomena and controlling the onset of turbulence.

Simulations of nonlinear flow dynamics have demonstrated that the DTWs, designed in Part~1, can maintain laminar flow and achieve positive net efficiency. In contrast, the UTWs promote turbulence even with the initial conditions for which the uncontrolled flow remains laminar. Our analysis of the Reynolds-Orr equation shows that, compared to the uncontrolled flow, the DTWs (UTWs) reduce (increase) the production of kinetic energy. 

We have also examined the effects of DTWs on fully developed turbulent flows at low Reynolds numbers. It turns out that the DTWs with speed and frequency selected in Part~1 and large enough amplitudes can eliminate turbulence (i.e., relaminarize the flow). We also note that, in spite of promoting turbulence, the UTWs may still achieve smaller drag coefficients compared to the uncontrolled flow. By increasing the UTW amplitude, even sub-laminar drag can be attained~\citep{minsunspekim06}. It is to be noted, however, that large wave amplitudes introduce poor net efficiency in flows subject to a fixed pressure gradient. Nevertheless, these traveling waves may still be utilized when the primal interest is to eliminate turbulence (with DTWs) or reduce the skin-friction drag (with UTWs) irrespective of the cost of control.


All simulations in the present study are enforced by a fixed pressure gradient, as opposed to the constant mass flux simulations of~\cite{minsunspekim06}. This is consistent with Part~1 where receptivity analysis was done for flows driven by a fixed pressure gradient. Even though these setups are equivalent in steady flows~\citep{scopio01}, they can exhibit fundamentally different behavior in unsteady flows. For example, the two simulations may possess different stability characteristics and yield structurally different solutions for near-wall turbulence~\citep{wal01,jimkawsim05}. Moreover,~\cite{jimkawsim05} remarked that the dynamical properties of the two simulations can significantly differ in small computational domains. Also,~\cite{ker05} suggested that specifying the simulation type comes next to defining the boundary conditions.

In order to examine the effect of simulation type on transition, skin-friction drag coefficient, and control net efficiency, we have repeated some of the simulations by adjusting the pressure gradient to maintain a constant mass flux. Our results reveal that regardless of the simulation type, the DTWs designed in Part~1 are effective in preventing transition while the UTWs promote turbulence. Moreover, the steady-state skin-friction drag coefficients are almost identical in both cases. However, the control net efficiency depends significantly on the simulation type. This is because of the difference in the definition of the produced power: in the fixed pressure gradient setup, the produced power is captured by the difference between the bulk fluxes in the uncontrolled and controlled flows; in the constant mass flux setup, the produced power is determined by the difference between the driving pressure gradients in the uncontrolled and controlled flows. It turns out that the produced power is larger in the constant mass flux simulation than in the fixed pressure gradient simulation, whereas the required power remains almost unchanged. Consequently, both DTWs and UTWs have larger efficiency in constant mass flux simulations. For example, in fixed pressure gradient setup of the present study, the UTWs with~$(c = -2,$~$\omega_x = 0.05$~$\alpha = \{0.05, 0.125\})$ have negative efficiency. The efficiency of these UTWs is positive, however, in constant mass flux simulations~\citep{minsunspekim06}. Our ongoing effort is directed towards understanding the reason behind this disagreement which may be ultimately related to the fundamental difference between these two types of simulations.

\section*{Acknowledgements}
   \label{sec.Ack}

    Financial support from the National Science Foundation under CAREER Award CMMI-06-44793 and 3M Science and Technology Fellowship (to R.\ M.) is gratefully acknowledged. The authors also thank Dr.\ John Gibson for helpful discussions regarding the Channelflow code. The University of Minnesota Supercomputing Institute is acknowledged for providing computing resources.

\end{document}

%% file: commands.tex
\newcommand{\beq}{\begin{equation}}
\newcommand{\eeq}{\end{equation}}
\newcommand{\bseq}{\begin{subequations}}
\newcommand{\eseq}{\end{subequations}}
\newcommand{\beqn}{\begin{eqnarray}}
\newcommand{\eeqn}{\end{eqnarray}}
\newcommand{\ba}{\begin{array}}
\newcommand{\ea}{\end{array}}
\newcommand{\bct}{\begin{center}}
\newcommand{\ect}{\end{center}}
\newcommand{\btmz}{\begin{itemize}}
\newcommand{\etmz}{\end{itemize}}
\newcommand{\benum}{\begin{enumerate}}
\newcommand{\eenum}{\end{enumerate}}

\newcommand{\bv}{{\bf v}}

\newcommand{\be}{\begin{equation}}
\newcommand{\ee}{\end{equation}}

\newcommand{\cplxs}{ C\kern -.35em \rule{0.03 em}{.7 ex}~   }

\def\complex{\hbox{C\kern -.45em \rule{0.03 em}{1.5 ex}}~}

\newcommand{\bi}{\begin{itemize}}
\newcommand{\ei}{\end{itemize}}





%% file: arxiv-tw-Part2-06-23-10.bbl
\begin{thebibliography}{22}
\expandafter\ifx\csname natexlab\endcsname\relax\def\natexlab#1{#1}\fi

\bibitem[Bewley(2009)]{bew09}
{\sc Bewley, T.~R.} 2009 A fundamental limit on the balance of power in a
  transpiration-controlled channel flow. {\em J. Fluid Mech.\/} {\bf 632},
  443--446.

\bibitem[Borrero-Echeverry {\em et~al.\/}(2010)Borrero-Echeverry, Schatz \&
  Tagg]{borschtag10}
{\sc Borrero-Echeverry, D., Schatz, M.~F. \& Tagg, R.} 2010 Transient
  turbulence in {T}aylor-{C}ouette flow. {\em Phys. Rev. E\/} {\bf 81}~(2),
  25301.

\bibitem[Brosa(1989)]{bro89}
{\sc Brosa, U.} 1989 Turbulence without strange attractor. {\em J. Stat.
  Phys.\/} {\bf 55}~(5), 1303--1312.

\bibitem[Canuto {\em et~al.\/}(1988)Canuto, Hussaini, Quarteroni \&
  Zang]{canhusquazan88}
{\sc Canuto, C., Hussaini, M.~Y., Quarteroni, A. \& Zang, T.~A.} 1988 {\em
  Spectral Methods in Fluid Dynamics\/}. New York, NY: Springer-Verlag.

\bibitem[Currie(2003)]{cur03}
{\sc Currie, I.~G.} 2003 {\em Fundamental Mechanics of Fluids\/}. CRC Press.

\bibitem[Fukagata {\em et~al.\/}(2009)Fukagata, Sugiyama \&
  Kasagi]{fuksugkas09}
{\sc Fukagata, K., Sugiyama, K. \& Kasagi, N.} 2009 {On the lower bound of net
  driving power in controlled duct flows}. {\em Physica D: Nonlinear
  Phenomena\/} {\bf 238}~(13), 1082--1086.

\bibitem[Gibson(2007)]{gib07}
{\sc Gibson, J.~F.} 2007 Channelflow: a spectral {N}avier-{S}tokes solver in
  {C}{++}. {\em Tech. Rep.\/} {G}eorgia {I}nstitute of {T}echnology,
  \url{www.channelflow.org}.

\bibitem[Grossmann(2000)]{gro00}
{\sc Grossmann, S.} 2000 The onset of shear flow turbulence. {\em Rev. Mod.
  Phys.\/} {\bf 72}, 603--618.

\bibitem[H{\oe}pffner \& Fukagata(2009)]{hoefuk09}
{\sc H{\oe}pffner, J. \& Fukagata, K.} 2009 Pumping or drag reduction? {\em J.
  Fluid Mech.\/} {\bf 635}, 171--187.

\bibitem[Hof {\em et~al.\/}(2006)Hof, Westerweel, Schneider \&
  Eckhardt]{hofwesscheck06}
{\sc Hof, B., Westerweel, J., Schneider, T.~M. \& Eckhardt, B.} 2006 Finite
  lifetime of turbulence in shear flows. {\em Nature\/} {\bf 443}~(7), 59 --
  62.

\bibitem[Jim\'{e}nez {\em et~al.\/}(2005)Jim\'{e}nez, Kawahara, Simens, Nagata
  \& Shiba]{jimkawsim05}
{\sc Jim\'{e}nez, J., Kawahara, G., Simens, M.~P., Nagata, M. \& Shiba, M.}
  2005 Characterization of near-wall turbulence in terms of equilibrium and
  ``bursting'' solutions. {\em Phys. Fluids\/} {\bf 17}~(1), 015105.

\bibitem[Jovanovi\'{c}(2008)]{jovPOF08}
{\sc Jovanovi\'{c}, M.~R.} 2008 Turbulence suppression in channel flows by
  small amplitude transverse wall oscillations. {\em Physics of Fluids\/} {\bf
  20}, 014101.

\bibitem[Jovanovi\'c {\em et~al.\/}(2006)Jovanovi\'c, Moarref \&
  You]{jovmoayouCTR06}
{\sc Jovanovi\'c, M.~R., Moarref, R. \& You, D.} 2006 Turbulence suppression in
  channel flows by means of a streamwise traveling wave. In {\em Proceedings of
  the 2006 Summer Program\/}, pp. 481--494. Center for Turbulence Research,
  Stanford University/NASA.

\bibitem[Kerswell(2005)]{ker05}
{\sc Kerswell, R.~R.} 2005 Recent progress in understanding the transition to
  turbulence in a pipe. {\em Nonlinearity\/} {\bf 18}, R17--R44.

\bibitem[Marusic {\em et~al.\/}(2007)Marusic, Joseph \& Mahesh]{marjosmah07}
{\sc Marusic, I., Joseph, D.~D. \& Mahesh, K.} 2007 Laminar and turbulent
  comparisons for channel flow and flow control. {\em J. Fluid Mech.\/} {\bf
  570}, 467--477.

\bibitem[Min {\em et~al.\/}(2006)Min, Kang, Speyer \& Kim]{minsunspekim06}
{\sc Min, T., Kang, S.~M., Speyer, J.~L. \& Kim, J.} 2006 Sustained sub-laminar
  drag in a fully developed channel flow. {\em J. Fluid Mech.\/} {\bf 558},
  309--318.

\bibitem[Moarref \& Jovanovi\'c(2010)]{moajov10}
{\sc Moarref, R. \& Jovanovi\'c, M.~R.} 2010 Controlling the onset of
  turbulence by streamwise traveling waves. {P}art 1: {R}eceptivity analysis.
  {\em J. Fluid Mech.\/} In press.

\bibitem[Peyret(2002)]{pey02}
{\sc Peyret, R.} 2002 {\em Spectral Methods for Incompressible Viscous Flow\/}.
  New York, NY: Springer.

\bibitem[Quadrio \& Ricco(2004)]{quaric04}
{\sc Quadrio, M. \& Ricco, P.} 2004 Critical assessment of turbulent drag
  reduction through spanwise wall oscillations. {\em J.\ Fluid Mech.\/} {\bf
  521}, 251--271.

\bibitem[Schmid \& Henningson(2001)]{schhen01}
{\sc Schmid, P.~J. \& Henningson, D.~S.} 2001 {\em Stability and Transition in
  Shear Flows\/}. Springer-Verlag.

\bibitem[Scotti \& Piomelli(2001)]{scopio01}
{\sc Scotti, A. \& Piomelli, U.} 2001 Numerical simulation of pulsating
  turbulent channel flow. {\em Phys. Fluids\/} {\bf 13}~(5), 1367--1384.

\bibitem[Waleffe(2001)]{wal01}
{\sc Waleffe, F.} 2001 Exact coherent structures in channel flow. {\em J.\
  Fluid Mech.\/} {\bf 435}, 93--102.

\end{thebibliography}
